\documentclass{aa}

\usepackage{graphicx}
\usepackage{txfonts}
\usepackage{mathabx}
\usepackage[normalem]{ulem}
\def \MJ{M$_{\mathrm{Jup}}$}
\def \MN{M$_{\mathrm{Nep}}$}
\def \ME{M$_{\Earth}$}
\def \RJ{R$_{\mathrm{Jup}}$}
\def \RN{R$_{\mathrm{Nep}}$}
\def \RE{R$_{\Earth}$}
\def \RS{R$_{\odot}$}
\def \msol{M$\mathrm{_\odot}$}

\def \1s{$1\,\sigma$}

\def \t0{T$_0$}

\def \aumic{AU~Mic}
\def \aumicb{AU~Mic~b}
\def \aumicc{AU~Mic~c}

\begin{document}

   \title{New constraints on the planetary system around the young active star AU Mic}
   \subtitle{Two transiting warm Neptunes near mean-motion resonance}
   \titlerunning{New constraints on the young planetary system AU Mic}

   \author{ E. Martioli
            \inst{1,2} \and
    G. H\'ebrard \inst{1,3} \and
    A. C. M. Correia \inst{4,5} \and
    J. Laskar \inst{5} \and
    A. Lecavelier des Etangs \inst{1}
    }

   \institute{
   \inst{1} Institut d’Astrophysique de Paris, CNRS, UMR 7095, Sorbonne Université, 98 bis bd Arago, 75014 Paris, France,
   \email{martioli@iap.fr} \\
   \inst{2} Laborat\'{o}rio Nacional de Astrof\'{i}sica, Rua Estados Unidos 154, 37504-364, Itajub\'{a} - MG, Brazil\\
   \inst{3} Observatoire de Haute Provence, St Michel l'Observatoire, France \\
   \inst{4} CFisUC, Department of Physics, University of Coimbra, 3004-516 Coimbra, Portugal \\
   \inst{5} IMCCE, UMR8028 CNRS, Observatoire de Paris, PSL University, Sorbonne Univ., 77 av. Denfert-Rochereau, 75014 Paris, France
    }

   \date{Received December 24, 2020; accepted March 27, 2021}

  \abstract {
  AU Microscopii
(\aumic) is a young, active star whose transiting planet was recently detected. Here, we report our analysis of its TESS light curve, where we modeled the BY Draconis type quasi-periodic rotational modulation by starspots simultaneously to the flaring activity and planetary transits. We measured a flare occurrence rate in \aumic\ of 6.35 flares per day for flares with amplitudes in the range of $0.06\% < f_{\rm max} < 1.5\%$ of the star flux. We employed a Bayesian MCMC analysis to model the five transits of \aumicb\ observed by TESS, improving the constraints on the planetary parameters. The measured planet-to-star effective radius ratio of $R_{\rm p}/R_{\star}=0.0496\pm0.0007$ implies a physical radius of $4.07\pm0.17$~\RE\ and a planet density of $1.4\pm0.4$~g\,cm$^{-3}$, confirming that \aumicb\ is a Neptune-size moderately inflated planet. While a single feature possibly due to a second planet was previously reported in the former TESS data, we report the detection of two additional transit-like events in the new TESS observations of July 2020. This represents substantial evidence for a second planet (\aumicc) in the system. We analyzed its three available transits and obtained an orbital period of $18.859019\pm0.000016$~d and a planetary radius of $3.24\pm0.16$~\RE, which defines \aumicc\ as a warm Neptune-size planet with an expected mass in the range of 2.2~\ME$< M_{\rm c} < $25.0~\ME, estimated from the population of exoplanets of similar sizes. The two planets in the \aumic\ system are in near 9:4 mean-motion resonance. We show that this configuration is dynamically stable and should produce transit-timing variations (TTV). Our non-detection of significant TTV in \aumicb\ suggests an upper limit for the mass of \aumicc\ of $<7$~\ME, indicating that this planet is also likely to be inflated.  As a young multi-planet system with at least two transiting planets, \aumic\ becomes a key system for the study of atmospheres of infant planets and of planet-planet and planet-disk dynamics at the early stages of planetary evolution.
  }


   \maketitle
%
\section{Introduction}

Young planetary systems represent an opportunity to observe planets in the early stages of planetary formation when gravitational interactions have not significantly changed the initial configuration of the system.  AU Microscopii (\aumic) is a particularly interesting young system, with an estimated age of $22\pm3$~Myr \citep{mamjek2014}, which is located at a distance of only $9.7248\pm0.0046$~pc \citep{gaiadr22018}. Table \ref{tab:aumicstarparams} summarizes the star parameters of \aumic\ that are relevant for this work. The \aumic\ host is an M1 star with a spatially resolved edge-on debris disk \citep{kalas2004} and at least one transiting planet, \aumicb\ \citep{plavchan2020Natur}. This Neptune-size planet is in a $8.5$-d prograde orbit aligned with the stellar rotation axis \citep{martioli2020, hirano2020, palle2020}. While \cite{plavchan2020Natur} only reported an upper limit on the mass of \aumicb, \cite{klein2020} measured $17.1^{+4.7}_{-4.5}$\ME\ thanks to infrared observations secured with the SPIRou spectropolarimeter \citep{donati2020}. \cite{plavchan2020Natur} has also reported the detection of an isolated transit event of a second possible candidate planet, hereafter referred to as \aumicc.

The notion that \aumic\ could be a host for several planets is not a surprising one. Indeed, the occurrence rate of planets with a radius between 0.5~R$_{\Earth}$ and 4.0~R$_{\Earth}$ and a period between 0.5 and 256~d is estimated as $8.4^{+1.2}_{-1.1}$ planets per M dwarf \citep{Hsu2020}, although the occurrence rate of more massive planets in M dwarfs decreases significantly with mass \citep[e.g.,][]{bonfils2013}. Moreover, provided the fact that \aumic\ has an edge-on debris disk and that \aumicb\ is a transiting planet with an aligned orbit, the chances that other planets also reside in co-planar orbits increase and, therefore, possible close-in additional planets are also likely to transit the star.

\aumic\ is a magnetically active star with strong flaring activity \citep{Robinson2001}. Its surface is largely filled by starspots, producing a BY Draconis-type light curve with a quasi-periodic rotational modulation and a period of $4.863\pm0.010$~d \citep{plavchan2020Natur}.  As in other active stars, the strong flaring and magnetic activity of \aumic\ makes it more difficult to detect planetary transits in a photometric time series, especially for small planets where the occurrence rate is higher.  In this work, we present an analysis of the Transiting Exoplanet Survey Satellite (TESS) data of \aumic, including the new observations obtained in July 2020, where we implement a multi-flare model combined with the starspots model, improving the constraints on the planetary parameters from transit modeling and allowing for the detection of two additional transits of the second candidate planet \aumicc.

\begin{table}
\centering
\caption{\aumic\ star parameters.}
\label{tab:aumicstarparams}
\begin{tabular}{lcc}
\hline
Parameter & Value & Ref. \\
\hline
effective temperature & $3700\pm100$~K & 1 \\
star mass  & $0.50\pm0.03$~\msol & 1 \\
star radius  & $0.75\pm0.03$~\RS & 2 \\
rotation period & $4.863\pm0.010$~d & 1 \\
age  & $22\pm3$~Myr & 3 \\
distance  & $9.7248\pm0.0046$~parsec & 4 \\
linear limb dark. coef.  & 0.2348 & 5 \\
quadratic limb dark. coef.  & 0.3750 & 5 \\
\hline
\end{tabular}
\tablebib{
(1) \citet{plavchan2020Natur};
(2) \citet{russel2015};
(3) \citet{mamjek2014};
(4) \citet{gaiadr22018};
(5) \citet{claret2018}
}
\end{table}

\section{TESS light curve}
\label{sec:tesslightcurve}

\begin{figure*}
   \centering
   \includegraphics[width=1\hsize]{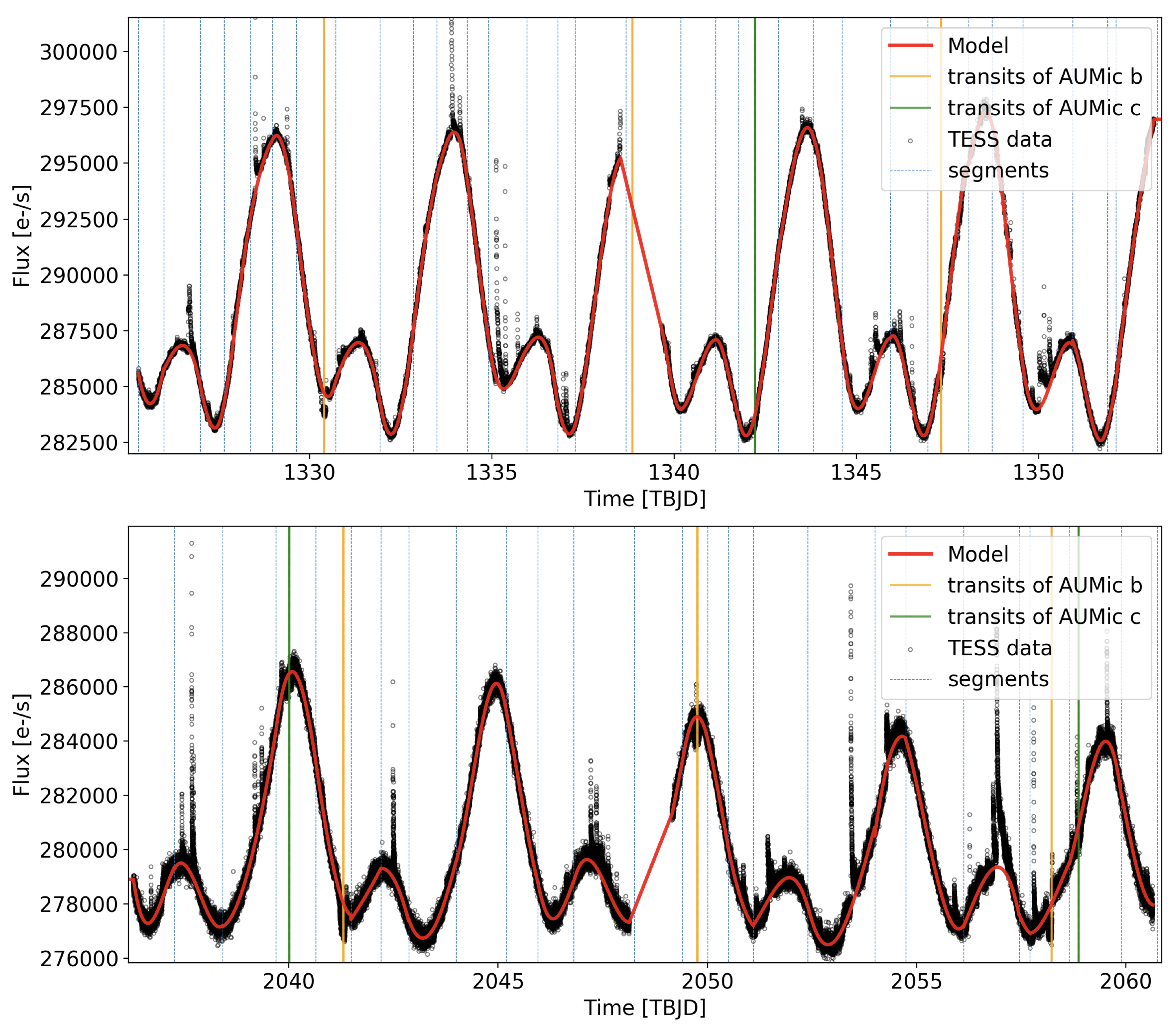}
      \caption{TESS light curve of \aumic. Top panel:\ Data for the first visit (from 2018-Jul-25 to 2018-Aug-22). Bottom panel: Data for the second visit (from 2020-Jul-04 to 2020-Jul-30). Black points are the PDCSAP flux and vertical solid lines show the times of planetary transits, as obtained in our analysis, for \aumicb\ (in orange) and \aumicc\ (in green). Blue vertical dashed lines show the knots of selected ranges for the piece-wise polynomial fit and the red line shows the starspot model.
    }
        \label{fig:tesslc}
  \end{figure*}

\aumic\ was observed by TESS \citep{tess_paper} in Sector 1 from 2018-Jul-25 to 2018-Aug-22, in cycle 1, camera 1. Then it was observed again in Sector 27 from 2020-Jul-04 to 2020-Jul-30, in cycle 3, camera 1. The second visit was observed as part of the TESS Guest Investigator Programs G03263 (PI: P. Plavchan), G03141 (PI: E. Newton), and G03273 (PI: L. Vega) in fast mode with a time sampling of 20 sec compared to the 2 min time sampling of the first visit. We obtained the de-trended photometric time series from the Mikulski Archive for Space Telescopes (MAST) using the MAST \texttt{astroquery} tool \footnote{\url{https://astroquery.readthedocs.io/en/latest/mast/mast.html}}.  Figure~\ref{fig:tesslc} shows the \aumic\ PDCSAP\footnote{Pre-search Data Conditioning SAP} flux in units of electrons-per-second (e$^{-}$s$^{-1}$) as a function of time given in TESS Barycentric Julian Date (TBJD = BJD - 2,457,000.0) for the two visits.  \citet{plavchan2020Natur} have analyzed the same TESS data from the first visit only, and in the present paper we report for the first time an analysis including the more recent 2020 TESS data. The predicted times of the transits of \aumicb\ and c as calculated from our ephemerides (see Sects. \ref{sec:aumicbtransits} and \ref{sec:aumicc}) are marked with vertical lines. 

\section{Starspots}
\label{sec:starspots}

As illustrated in Fig.~\ref{fig:tesslc}, the TESS light curve of \aumic\ shows a typical BY Draconis type of quasi-periodic variation due to starspots modulated by stellar rotation.  The starspots evolve and, therefore, this modulation cannot be accurately modeled by a single periodic function. Thus, we treat this quasi-periodic modulation as a baseline ``continuum'' signal, where we model it by fitting a piece-wise fourth-order polynomial with iterative sigma-clipping. This approach works well when the ranges are carefully selected and inspected as follows. First, we can be certain that the polynomial function is a sufficiently accurate approximation for the local baseline variation within each range. Then we avoid the edges of the ranges to include either a transit or a flare event. Finally, we avoid the inclusion of large gaps (when TESS did not deliver data) within the same range. The edges for all selected ranges are represented by the blue vertical dashed lines in Fig.~\ref{fig:tesslc}.

This starspot model is removed from the data to model flares and planetary transits as explained in Sects. \ref{sec:flares}, \ref{sec:aumicbtransits}, and \ref{sec:aumicc}, which are further removed from the original data to fit the starspot modulation again. We repeat this procedure iteratively until the standard deviation of residuals is not improved by more than 1\%. We typically meet this criterion upon three to five iterations. The final starspot model is also shown in Fig.~\ref{fig:tesslc}. In Appendix \ref{app:starrotation}, we present an independent measurement of the rotational period of \aumic\ from the starspot modulation in the 2018 TESS data.

\section{Flares}
\label{sec:flares}

\begin{figure*}
   \centering
   \includegraphics[width=1\hsize]{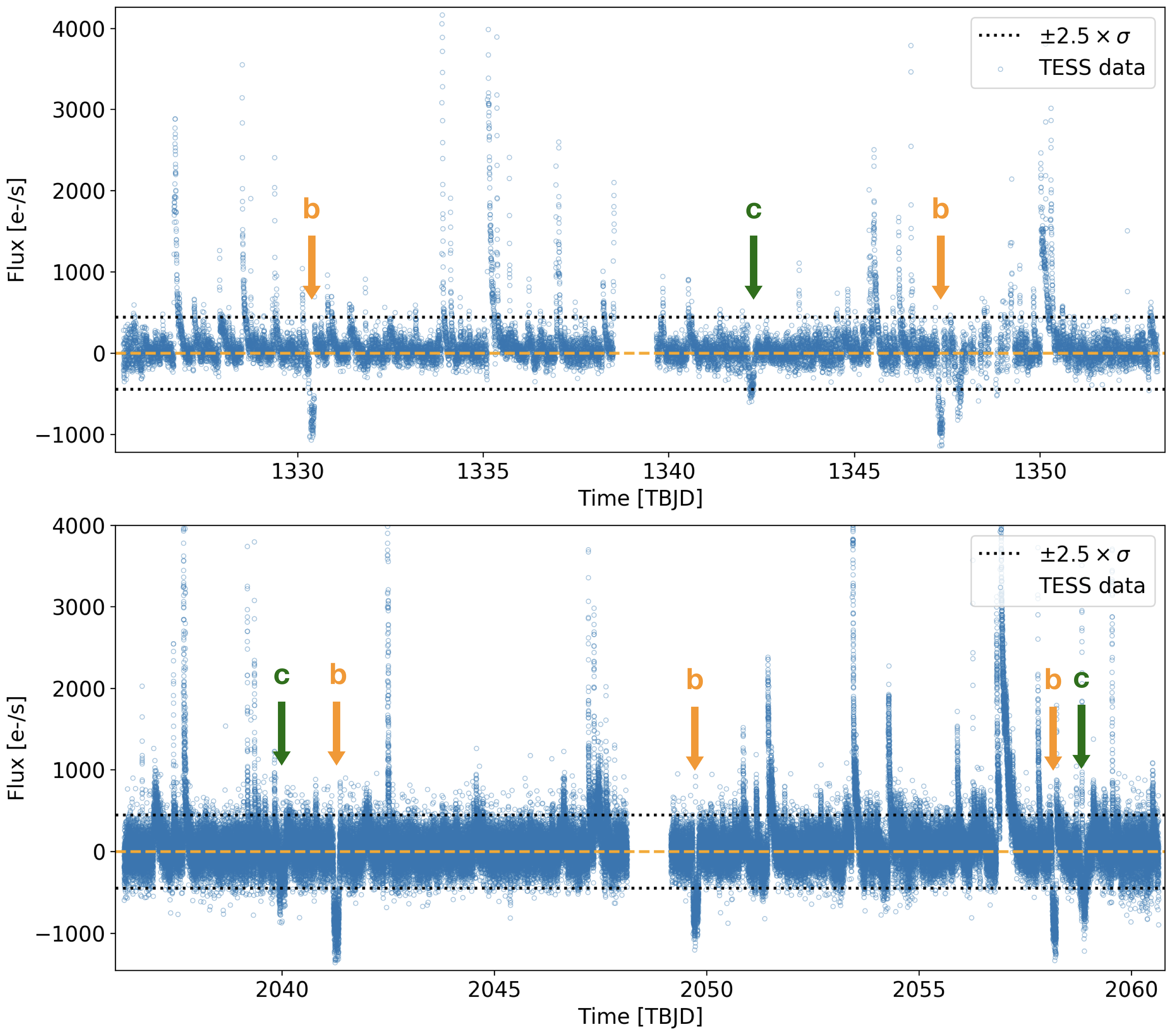}
      \caption{TESS residual light curve of \aumic. Top panel: Residual light curve for the first TESS visit (2018-Jul-25 to 2018-Aug-22). Bottom panel: Residual curve for the second visit (2020-Jul-04 to 2020-Jul-30). Blue circles show \aumic\ TESS flux data after subtracting the starspot model and black dotted lines show $\pm 2.5\sigma$ range around the starspot fit model, for $\sigma=162.5$~e$^{-}$s$^{-1}$. The flaring activity can be seen as positive groups of points above $+2.5\sigma$ and the transits as negative groups of points below $-2.5\sigma$. Our predicted times of transits for \aumicb\ (in orange) and \aumicc\ (in green) are marked with arrows.
              }
        \label{fig:tess-cont-sub-lc}
  \end{figure*}

 \aumic\ is an active young star with intense flaring activity. We carry out an analysis of the flares in the TESS data to improve the constraints on the transit events. We consider the starspot-subtracted residuals as shown in Fig.~\ref{fig:tess-cont-sub-lc}. The flares are typically clusters of points lying above the noise level. First we detect peaks as possible candidate flares using the \texttt{scipy.signal.find\_peaks}\footnote{\url{https://docs.scipy.org/doc/scipy/reference/generated/scipy.signal.find_peaks.html}} routine, where it finds local maxima via a simple comparison of neighbouring values. We set a minimum peak amplitude of $2.5\sigma$ and a minimum horizontal distance between peaks of about $\sim100$ minutes. Some flares in \aumic\ are quite complex, while other flares occur before a previously initiated flare has ended. Therefore, we visually inspected  all detected peaks to identify any possible additional peaks that have been missed by the find-peaks algorithm. After a few iterations, we identified a total of 324 flares in the whole TESS time series.

\begin{figure}
   \centering
   \includegraphics[width=1\hsize]{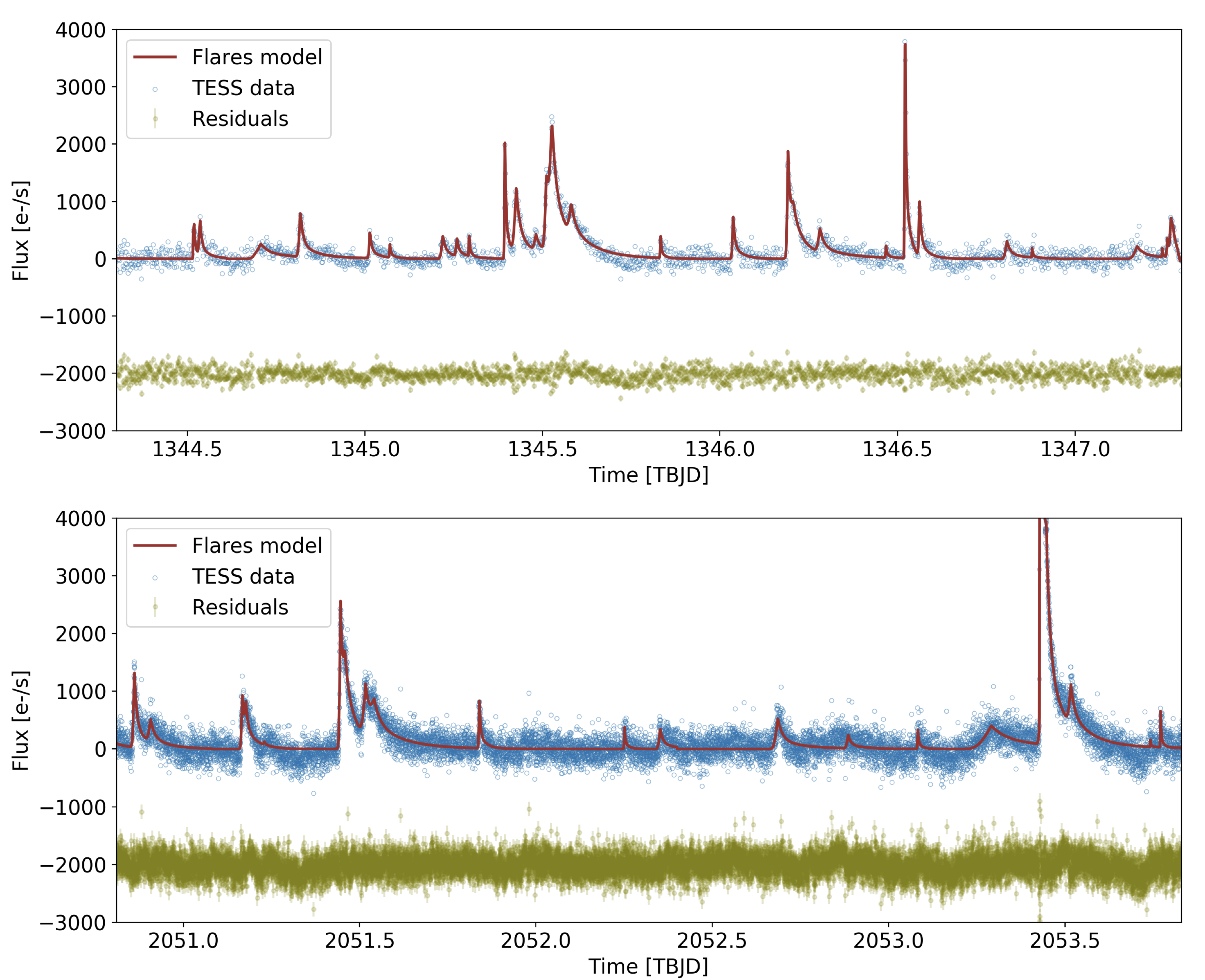}
      \caption{TESS residual light curve of \aumic\ between TBJD=1344.3 and TBJD=1347.3 (top panel) and between TBJD=2050.8 and TBJD=2053.8 (bottom panel). Blue circles show the flux after subtracting the starspot model, brown line shows the multi-flare best fit model. Green points show the residuals with an arbitrary offset for better visualization. The dispersion of residuals are 107~e$^{-}$s$^{-1}$ and 178~e$^{-}$s$^{-1}$ for data presented in top panel and bottom panel, respectively.
      }
    \label{fig:flares}
  \end{figure}

 The times and amplitudes of detected peaks were adopted as initial values for  the basis of performing a least-squares fit using the multi-flare model of \cite{Davenport2014}, where each flare is represented by a two-phase model with a polynomial rise and an exponential decay. Figure~\ref{fig:flares} shows, as an example, the residual light curve and fit flares model for three days of flaring activity in \aumic\ in both TESS visits. For each flare in the model, we fit the flare amplitude ($f_{\rm max}$), the full width at half maximum (FWHM), and the time of maximum $t_{\rm peak}$. The fit parameters for all flares are presented in Appendix \ref{app:flaresfitparams}.

\begin{figure}
   \centering
   \includegraphics[width=1\hsize]{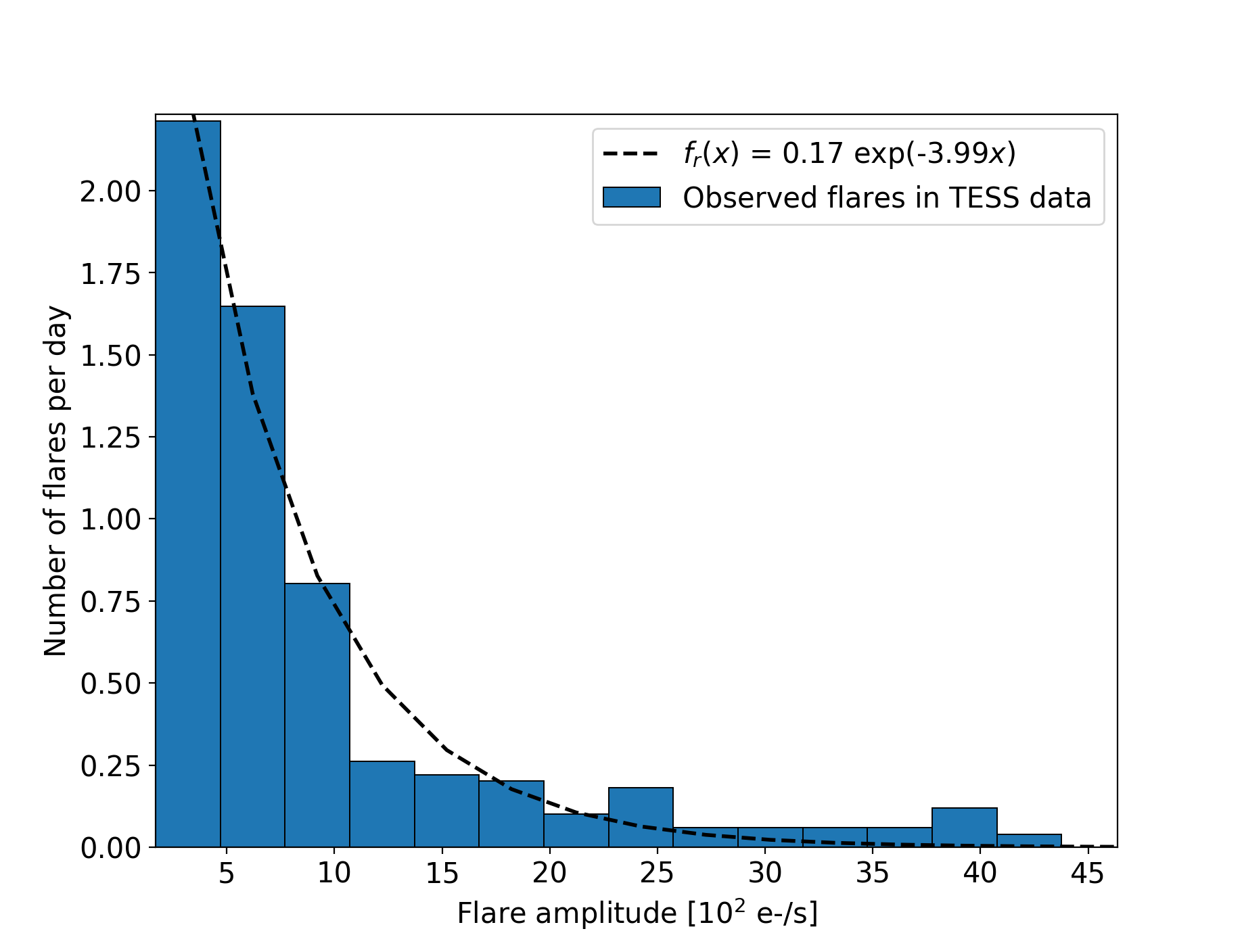}
      \caption{Number of flares normalized by the total observed time of 49.75~d as function of flare amplitude. Dashed line shows the fit to an exponential decay as shown in the legend, where variable $x$ represents the flare amplitude.
            }
        \label{fig:flarerate_amplitude}
  \end{figure}

 The total continuous monitoring time of \aumic\ performed by TESS was estimated as 49.75~d, where we have discounted the large gaps in the data. We detected a total of 324 flares, from which 316 have fit amplitudes above $F_{\rm max}>1\sigma$, for $\sigma=161.5$~e$^{-}$s$^{-1}$.   Thus, the \aumic\ global occurrence rate of flares with amplitude above $f_{\rm max}\gtrsim0.06$\% of the median stellar flux ($279544$~e$^{-}$s$^{-1}$) is estimated as 6.35 flares per day or about 1 flare every 3.8 hours.  Figure~\ref{fig:flarerate_amplitude} shows the distribution of flares as function of flare amplitudes. We fit an exponential to the measured distribution and obtained an empirical model for the occurrence rate as function of flare amplitude of $f_r = 0.17{\rm e}^{-4.0 f_{\rm max}}$~d$^{-1}$, with $f_{\rm max}$ in units of $10^{2}$~e$^{-}$s$^{-1}$. We notice that the strong flares, which are more important energetically, have a much lower occurrence rate and therefore a low statistical significance in our analysis. Thus, this empirical relationship is only expected to hold  for flares above the detection limit imposed by the noise and for flares with a statistically significant number of events observed in the TESS data, that is, those with amplitudes in the range of $0.06\% < f_{\rm max} < 1.5\%$ of the star flux.

 The transit duration of \aumicb\ calculated by \cite{plavchan2020Natur} is about 3.5 hours with transit depth of 0.26\%. Therefore, considering that a transit observation of \aumicb\ requires an out-of-transit baseline of about 2~hours, it is most likely that any transit observation in \aumic\ should be affected in some way by flare events. This shows the importance of modeling flares adequately to improve the constraints on the planetary parameters obtained by the transits.

\section{Transits of \aumicb}
\label{sec:aumicbtransits}

Three transits of \aumicb\ occurred during the first TESS visit, but only two have been observed due to a communication problem with the spacecraft during the second transit \citep[see][]{plavchan2020Natur}. For the second visit, three other transits were expected, and indeed we were able to identify all of them. To analyze these five transits together, we start with the parameters of \aumicb\ from \cite{plavchan2020Natur}, referred to as "prior values" in Table \ref{tab:aumicbfitparams}, to remove the transit signal from the TESS light curve before fitting the starspot and flare models. Our transit model is calculated using the \texttt{BATMAN} toolkit by \cite{Kreidberg2015}, where we assume a circular orbit ($e=0$). This assumption is in agreement with \cite{plavchan2020Natur}, who did not detect any significant eccentricity from their analysis of the transits only, where the eccentricity would be slightly constrained by the duration of the transit. We adopt as priors the quadratic limb-darkening coefficients (LDC) from \cite{claret2018} with an arbitrary error of 0.1, where we obtained the values calculated for the photometric system of TESS and those matching the closest stellar parameters to those of \aumic\ (see Table \ref{tab:aumicstarparams}).

The transits of \aumicb, flares, and starspots are fit simultaneously using the non-linear least squares optimization (OLS) fit tool \texttt{scipy.optimize.leastsq}.  As explained in Sect. \ref{sec:starspots} this procedure is run iteratively, where we first obtain an independent fit for each component of the model and then we set these values as initial guess to run an OLS analysis with all free parameters in the three components of the model. Then we consider the data in the ranges around the transits of \aumicb\ to sample the posterior distributions of the transit parameters using the \texttt{emcee} Markov chain Monte Carlo (MCMC) package \citep{foreman2013}.  The chain is set with 100 walkers and 30000 MCMC steps of which we discard the first 5000. The MCMC samples are presented in Fig.~\ref{fig:aumic_b_mcmc_chain} in Appendix \ref{app:posteriordistributions}, which show the chains reaching stability before the first 5000 discarded steps. The posterior distributions are illustrated in Fig.~\ref{fig:aumic_b_pairsplot} in Appendix \ref{app:posteriordistributions}. The best-fit values of the transit parameters are calculated as the medians of the posterior distribution with error bars defined by the 34\% on each side of the median, all presented in Table \ref{tab:aumicbfitparams}.

\begin{figure*}
   \centering
       \includegraphics[width=1.0\hsize]{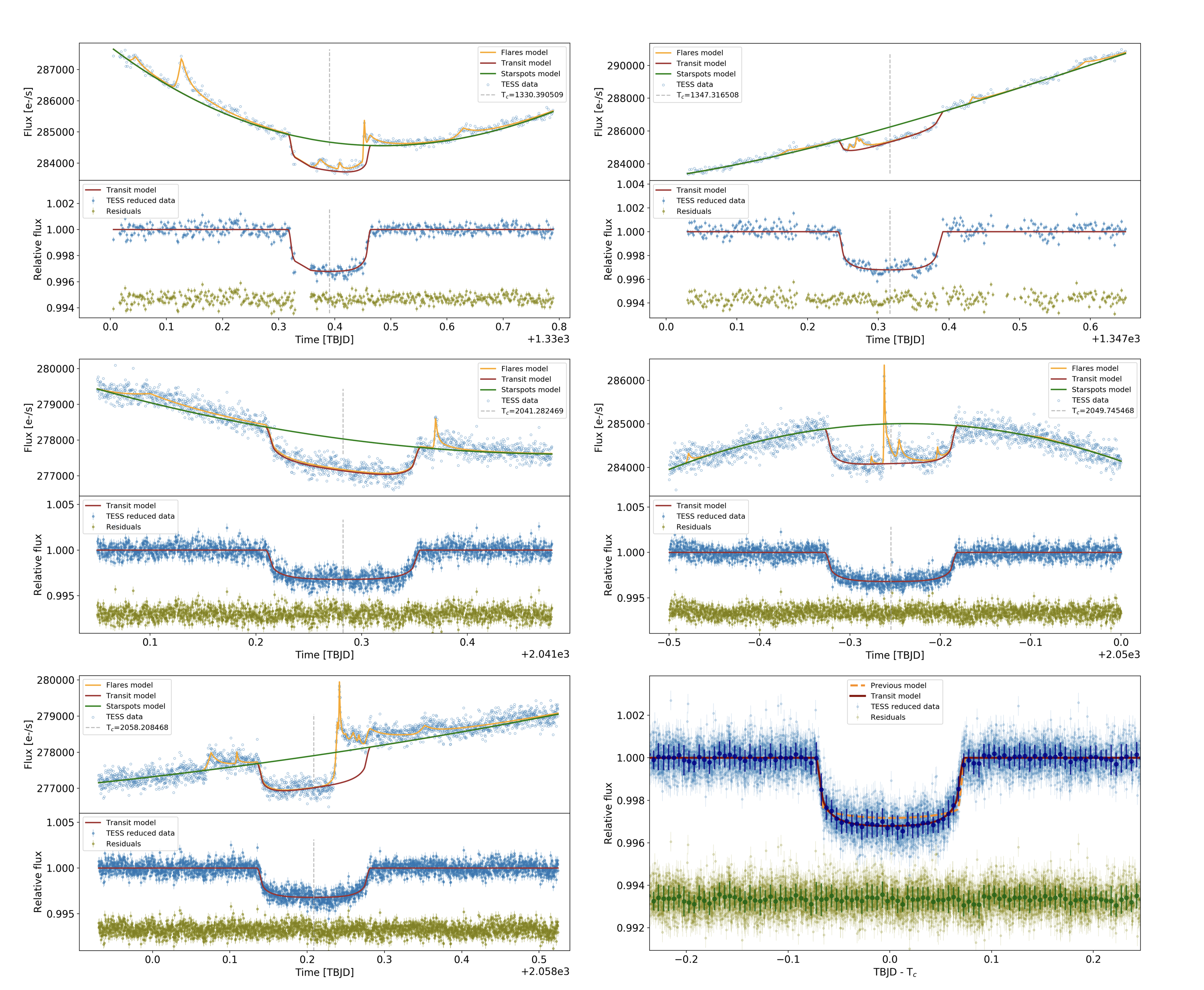}
      \caption{Each panel shows the range around one of the five transits of \aumicb\ observed by TESS. The top part of each panel shows the TESS data (blue circles) and the best-fit model for each component, as indicated in the legend. The bottom part in each panel shows the TESS data normalized by the starspot and flare model (blue points) and the best-fit transit model (red line). Vertical dashed lines show the central time of each transit. The bottom right panel shows the phase-folded light curve for all the data from the five transits (light blue points) and the binned data (dark blue points) with a bin size of 0.005~d, our best fit model (red line), and the previous fit model of \cite{plavchan2020Natur} for comparison (orange dashed line). The residuals (green points) are displayed with an arbitrary offset.
      }
        \label{fig:aumic_b_fit}
  \end{figure*}

Figure~\ref{fig:aumic_b_fit} shows the results of our analysis for each transit range separately, and the bottom right panel shows the flux normalized by the starspot and flare models for all transits together, and our best fit model along with the previous model of \cite{plavchan2020Natur} for comparison. The dispersion of residuals are 358, 433, 633, 566, and 588 ppm, for each respective epoch. The global dispersion is 573 ppm. Notice that our measured parameters of \aumicb\ agree within $3\sigma$ with the previous measurements by \cite{plavchan2020Natur} but with improved accuracy. Our measured planet-to-star radius ratio is slightly larger than that measured by \cite{plavchan2020Natur} (see lower, right panel of Fig.~\ref{fig:aumic_b_fit}). However, as described in Sect. \ref{sec:discussion}, our derived effective planetary radius of $1.05\pm0.04$~\RN\ is slightly smaller than the value $1.08\pm0.05$~\RN\ reported by \cite{plavchan2020Natur}.

  \begin{table*}
    \caption[]{Transit fit parameters for planet \aumicb\ obtained in our analysis from a simultaneous fit of the five transits observed by TESS as illustrated in Figs. \ref{fig:aumic_b_fit} and \ref{fig:aumic_b_pairsplot}. }
    \label{tab:aumicbfitparams}
    \begin{center}
    \begin{tabular}{lcc}
        \hline
        \noalign{\smallskip}
 Parameter & Prior value & MCMC fit value \\
        \noalign{\smallskip}
        \hline
 time of conjunction, $T_0$ [TBJD] & $1330.38957^{+0.00015}_{-0.00068}$  & $1330.39051\pm0.00015$\\
 orbital period, $P$ [d] & $8.46321\pm0.00004$ & $8.463000\pm0.000002$\\
 normalized semi-major axis, $a/R_{\star}$ &  $19.1^{+1.8}_{-1.6}$ & $19.1^{+0.2}_{-0.4}$\\
 orbital inclination, $i$ [degree] & $89.5^\pm0.4$& $89.5^{+0.4}_{-0.3}$\\
 planet-to-star radius ratio, $R_{\rm p}/R_{\star}$ & $0.0514\pm0.0013$& $0.0526^{+0.0003}_{-0.0002}$\\
 linear limb dark. coef., $u_{0}$ & $\mathcal{N}(0.23,0.1)$ & $0.13\pm0.03$\\
 quadratic limb dark. coef., $u_{1}$ & $\mathcal{N}(0.38,0.1)$ & $0.58\pm0.06$\\
        \hline
    \end{tabular}
    \end{center}
  \end{table*}

\section{Confirmation and characterization of planet \aumicc}
\label{sec:aumicc}

An isolated candidate transit event at $T_{c}=1342.22\pm0.03$~TBJD was identified in the 2018 TESS data by \cite{plavchan2020Natur}, which was interpreted as a possible second planet in a $30\pm6$-d orbit. We have searched for other transit-like events in the TESS data by looking at regions with systematic flux dips below the noise level.  We identified three regions as potential transits including the one previously identified by \cite{plavchan2020Natur} (see Fig.~\ref{fig:tess-cont-sub-lc}). We notice that there is an additional region presenting a significant flux dip around ${\rm TBJD}=1347.8$, which did not include as a possible transit. The 2018 TESS data after ${\rm TBJD}\sim1347.7$ contain several gaps, which have an impact on the modeling of the baseline modulation by starspots as well as on the detection and modeling of flares. Therefore, we presume that the apparent flux dip in this region is more likely due to a misfit other than a transit. In addition the shape of that dip is different from that of the three features above.

\begin{figure*}
   \centering
       \includegraphics[width=1.0\hsize]{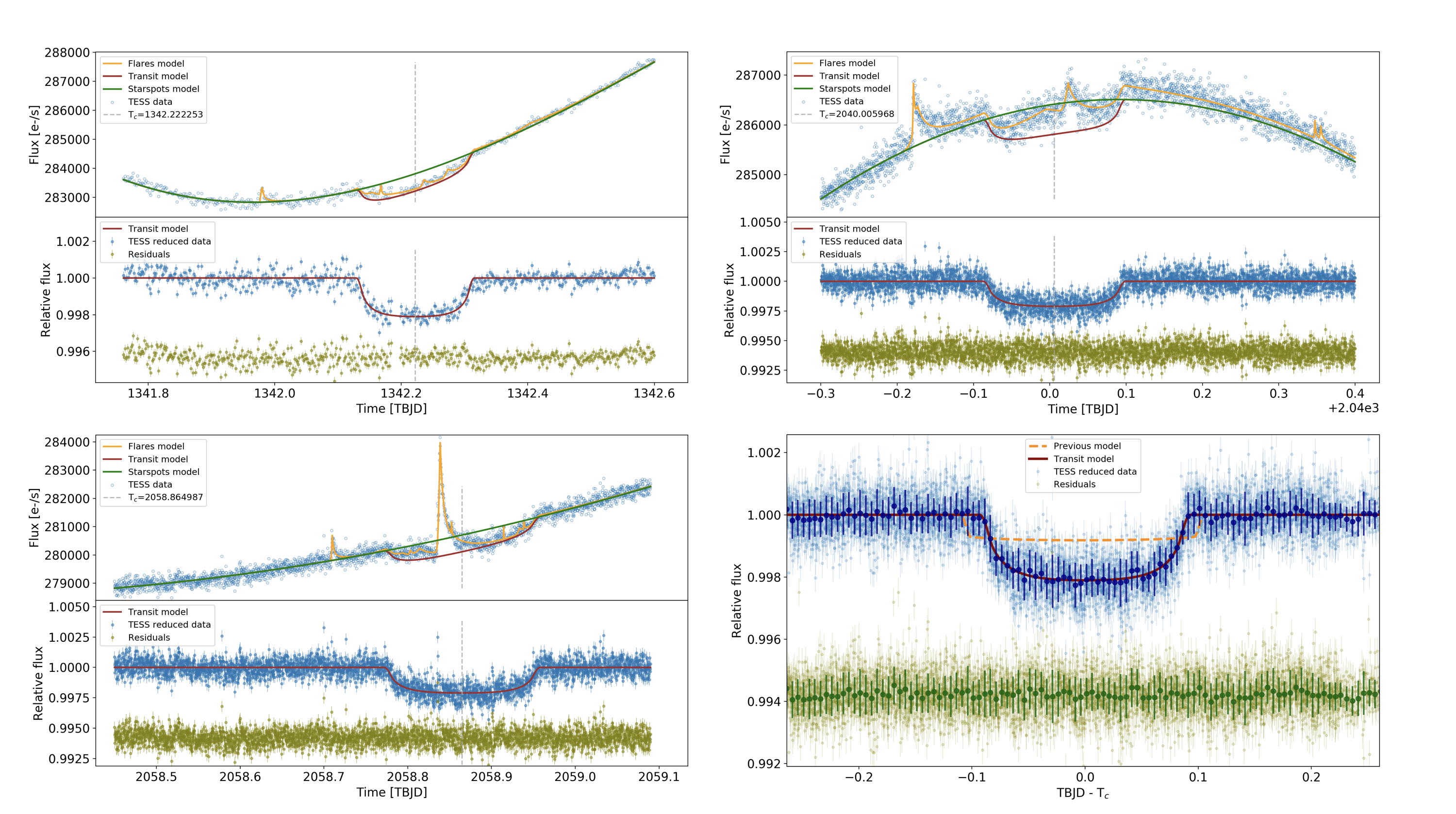}
      \caption{Each panel shows the range around one of the five transits
of \aumicb\ observed by TESS. Details are the same as in Fig.~\ref{fig:aumic_b_fit}, but here we present the results for the three TESS transits of \aumicc. The dispersion of residuals are 370, 646, 609 ppm, for each respective epoch. The global dispersion is 609 ppm.}
        \label{fig:aumic_c_fit}
  \end{figure*}

We hypothesize that the three selected events are transits of the same candidate planet \aumicc.  In order to test this hypothesis, we first obtain an orbital period that is consistent with the three transits, that is, a period given by the slope of a linear fit to the ephemeris equation, $T_{c} = T_{0} + P \times E$, with $T_{c}$ as the times of conjunction measured independently for each event and $E$ as the corresponding epochs to each event, assumed to be 0, 37, and 38. We find an orbital period of $P=18.85895\pm0.00003$~d.  However, we note that with this particular periodicity, only three events would occur during the TESS observations (see Fig.~\ref{fig:tesslc}). The possibility of shorter alias periods is ruled out by the fact that TESS would have observed more transits of this object and we did not detect them.
To test further our hypothesis, we performed an independent analysis of each event assuming an orbital period, first with an uniform prior of $P=\mathcal{U}(1,500)$~d to explore a broad range of periods, and then with a normal prior of $P=\mathcal{N}(18.86,0.01)$~d, which explores a solution constrained by the previous knowledge imposed by our hypothesis.  Analogously, the normalized semi-major axes are also first set with an uniform prior of $a/R_{\star}=\mathcal{U}(1,500)$ and then with a normal prior estimated using the Kepler's law, that is, $a/R_{\star}=\mathcal{N}(31.5,1.4)$. The priors for the times of conjunction are measured locally on each event, for example, for the first event, we obtained $T_{0}=\mathcal{N}(1342.23,0.01)$. The prior for the planet-to-star radius ratio is estimated from the square root of the average flux depth of the three events, assuming a conservative error, that is, $R_{\rm p}/R_{\star}=\mathcal{N}(0.04,0.01)$. The orbital inclination is considered with an uniform prior distribution of $i=\mathcal{U}(85^{\circ},90^{\circ})$ and the limb darkening coefficients are fixed to the literature values presented in Table \ref{tab:aumicstarparams}. In Appendix \ref{app:individualtransitsofaumicc}, we present the results for this independent analysis of each transit. The posterior of all the transit parameters agree within $3\sigma$ (see Table \ref{tab:othertransitsfitparams}), and the dispersion of residuals improves when we adopted the fit period of $P=\mathcal{N}(18.86,0.01)$~d as prior, which supports our hypothesis that these three events were caused by the transits of the same planet \aumicc.
Finally, we perform a joint analysis of the three transits simultaneously, where we adopt the same priors as in the independent analysis above, except for the limb darkening coefficients, where we adopted a normal prior with the literature values and an arbitrary error of 1.0. We call attention to the fact that the limb darkening coefficients obtained in Sect. \ref{sec:aumicbtransits} could have been used as priors since both planets are transiting the same star. However, the two planets may transit the stellar disk in different regions, and since \aumic\ is largely filled by starspots, the limb darkening may also be affected by the different temperatures of the transited regions in the photosphere.  Table \ref{tab:aumiccfitparams} presents the priors and fit parameters, while Fig.~\ref{fig:aumic_c_fit} shows the TESS data for the three transits and the best-fit model obtained from our analysis. As in Sect. \ref{sec:aumicbtransits}, the MCMC samples and posterior distributions are illustrated in Figs.~\ref{fig:aumic_c_mcmc_chain} and \ref{fig:aumic_c_pairsplot} in Appendix \ref{app:posteriordistributions}.

  \begin{table*}
    \caption[]{Transit fit parameters for planet \aumicc\ obtained in our analysis from a simultaneous fit of the three transits observed by TESS as illustrated in Figs.~\ref{fig:aumic_c_fit} and \ref{fig:aumic_c_pairsplot}. }
    \label{tab:aumiccfitparams}
    \begin{center}
    \begin{tabular}{lcc}
        \hline
        \noalign{\smallskip}
 Parameter & Prior value & MCMC fit value \\
        \noalign{\smallskip}
        \hline
 time of conjunction, $T_0$ [TBJD] & $\mathcal{N}(1342.23,0.01)$ & $1342.2223\pm0.0005$\\
 orbital period, $P$ [d] & $\mathcal{N}(18.86,0.01)$ & $18.859019^{+0.000016}_{-0.000015}$ \\
 normalized semi-major axis, $a/R_{\star}$ & $\mathcal{N}(31.5,1.4)$ & $29\pm3$ \\
 orbital inclination, $i$ [degree] & $\mathcal{U}(85,90)$ & $89.0^{+0.5}_{-0.4}$ \\
 planet-to-star radius ratio, $R_{\rm p}/R_{\star}$ & $\mathcal{N}(0.04,0.01)$ & $0.0418^{+0.0010}_{-0.0012}$ \\
 linear limb dark. coef., $u_{0}$ & $\mathcal{N}(0.23,1.0)$ & $0.0^{+0.2}_{-0.3}$ \\
 quadratic limb dark. coef., $u_{1}$ & $\mathcal{N}(0.38,1.0)$ & $1.2^{+0.4}_{-0.3}$ \\
        \hline
    \end{tabular}
    \end{center}
  \end{table*}

\section{Discussion}
\label{sec:discussion}

We combine the transit parameters of \aumicb\ and c presented in Tables \ref{tab:aumicbfitparams} and \ref{tab:aumiccfitparams} with the star parameters from Table \ref{tab:aumicstarparams} to calculate a number of derived parameters, which are presented in Table \ref{tab:derivedparams}.

  \begin{table*}
    \caption[]{Summary of the parameters of planets \aumicb\ and \aumicc. }
    \label{tab:derivedparams}
    \begin{center}
    \begin{tabular}{lccc}
        \hline
        \noalign{\smallskip}
Parameter & unit & \aumicb\ & \aumicc\ \\
        \noalign{\smallskip}
        \hline

time of conjunction & TBJD & $1330.39051\pm0.00015$ & $1342.2223\pm0.0005$ \\
orbital period & d &  $8.463000\pm0.000002$ &  $18.859019\pm0.000016$ \\

normalized semi-major axis, $a/R_{\star}$ & - & $19.1\pm0.3$ & $29\pm3$ \\
semi-major axis $^\dag$ & au & $0.0645\pm0.0013$ & $0.1101\pm0.0022$ \\

transit duration & h & $3.50\pm0.08$ & $4.5\pm0.8$ \\

orbital inclination & degree & $89.5\pm0.3$ & $89.0^{+0.5}_{-0.4}$\\
impact parameter & $R_{\star}$ & $0.18\pm0.11$ & $0.51\pm0.21$ \\

eccentricity & - & $0$ (FIXED) & $0$ (FIXED)\\

planet-to-star radius ratio~$^\S$, $R_{p}/R_{\star}$  & - & $0.0496\pm0.0007$ & $0.0395\pm0.0011$ \\
planet radius & \RJ & $0.371\pm0.016$ & $0.295\pm0.014$ \\
planet radius & \RN  & $1.05\pm0.04$ & $0.84\pm0.04$ \\
planet radius & \RE & $4.07\pm0.17$ & $3.24\pm0.16$ \\

velocity semi-amplitude & m\,s$^{-1}$ & $8.5^{+2.3}_{-2.2}$~$^{\S\S}$ & $0.8 < K_{\rm c} < 9.5$ \\

planet mass & \MJ & $0.054\pm0.015$~$^{\S\S}$ & 0.007 < M$_{\rm c}$ < 0.079\\
planet mass & \MN & $1.00\pm0.27$~$^{\S\S}$ &  0.13 < M$_{\rm c}$ < 1.46 \\
planet mass & \ME & $17\pm5$~$^{\S\S}$ & 2.2 < M$_{\rm c}$ < 25.0 \\

planet density & g\,cm$^{-3}$ & $1.4\pm0.4$ & $0.4 < \rho_{\rm c} < 4.1$ \\

equilibrium temperature & K & $593\pm21$ & $454\pm16$ \\
        \hline
    \end{tabular}

\tablebib{
$^\dag$ semi-major axis derived from the fit period and the Kepler's law; $^\S$ radius values correspond to the effective radius, which is already corrected for spot coverage and contains an additional uncertainty due to rotational modulation, as explained in Sect. \ref{sec:discussion}; $^{\S\S}$ \citet{klein2020}
}
    \end{center}
  \end{table*}

The planet-to-star radius ratio obtained in Sects. \ref{sec:aumicbtransits} and \ref{sec:aumicc} are likely overestimated due to the presence of spots covering a significant fraction of the stellar photosphere. This is because the observed transit depth is likely smaller than the nominal depth that would be obtained for a completely unspotted photosphere. Therefore, an effective planet-to-star radius ratio can be obtained by applying a correction factor to the observed transit depth as in Eq. 1 of \cite{Rackham2018}, that is,

\begin{equation}
\left(\frac{R_{p}}{R_{\star}}\right)_{\rm eff}^{2} = \left[1-f_{\rm spot}\left(1 - \frac{F_{\rm spot}}{F_{\rm phot}}\right)\right] \left(\frac{R_{p}}{R_{\star}}\right)_{\rm obs}^{2}
\end{equation}

for $f_{\rm spot}$ being the spot filling factor estimated at 20\% based on magnetic activity \citep{klein2020} and $F_{\rm spot}/F_{\rm phot}$ being the flux fraction between the spots and the photosphere. The latter can be estimated assuming a blackbody flux for each of the two components, where the photospheric temperature is $T_{\rm phot}=3700$~K and the spots' temperature is estimated at 86\% of the photospheric value in M1 stars \citep{Rackham2018}. The latter is in agreement with spectropolarimetric measurements of \aumic\ by \citet{Berdyugina2011}. Our estimation for the spot-to-photosphere flux fraction integrated over the TESS band pass (between 600~nm and 1000~nm) is $F_{\rm spot}/F_{\rm phot}\sim46$~\%. Therefore, the effective radius for both transiting planets in the \aumic\ system should be given by $R^{\rm eff}_{p} = 0.94 R^{\rm obs}_{p}$, that is, about 6\% smaller than the measured radius.

In addition to the correction above, there is an uncertainty in the planet radius due to the rotational modulation by starspots and other photospheric heterogeneities, which implies a variable baseline flux. To account for this additional uncertainty, we considered the amplitude of the photospheric modulation in the \aumic\ TESS light curve, which is about 5\%. This variability propagates an uncertainty on the order of $\sim \sqrt{0.05}$ to the planet radius measurements, which is included in the uncertainty values reported in Table \ref{tab:derivedparams}.

\cite{klein2020} recently measured a semi-amplitude of the radial velocity of the reflex motion caused by \aumicb\ of $K_{b}=8.5^{+2.3}_{-2.2}$~m\,s$^{-1}$, providing an important constraint on the mass of this planet. We recalculated the planet density based on our radius measurement and obtained $\rho_{b}=1.4\pm0.4$~g\,cm$^{-3}$.

\begin{figure*}
   \centering
       \includegraphics[width=1.0\hsize]{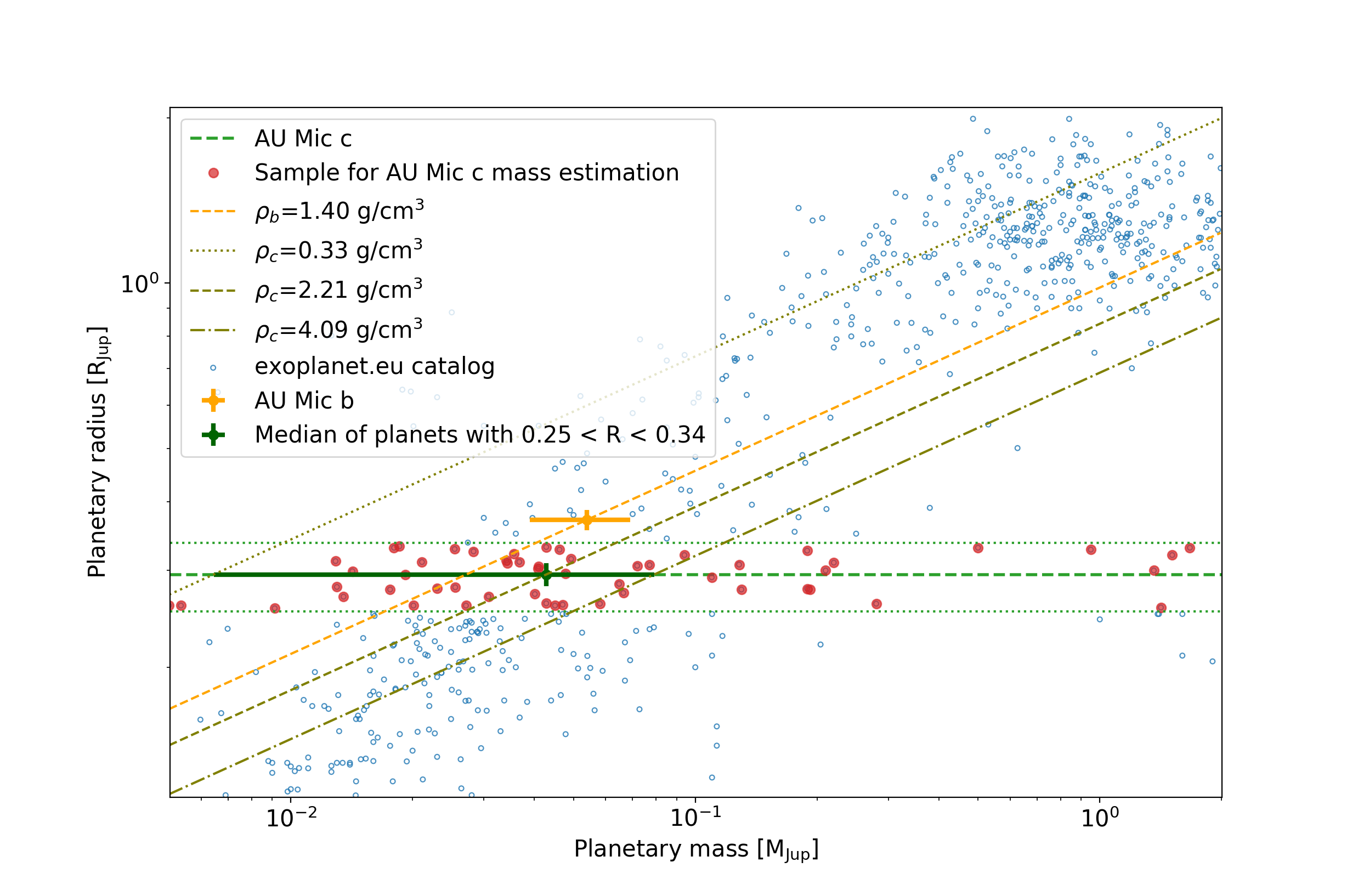}
      \caption{Mass-radius diagram of planets. Blue circles show the known transiting exoplanets compiled from {\tt exoplanet.eu}. The orange point represents \aumicb\ and the light green horizontal dashed and dotted lines represent the \aumicc\ radius measurement and $\pm1\sigma$ uncertainties. Dark green point represents the median of planets with radius within $1\sigma$ of the measured radius of \aumicc . Dashed orange line shows an iso-density line for the radius and mass of \aumicb\ and the olive lines correspond to the iso-densities obtained for the radius of \aumicc\ and for the lower limit (dotted), median value (dashed), and upper limit (dot-dashed) of the mass of \aumicc , based on the exoplanet population with the same radii.}
        \label{fig:mass_radius}
  \end{figure*}

To estimate a plausible mass range for \aumicc, we considered a population of transiting exoplanets \footnote{exoplanet parameters compiled from \url{exoplanet.eu}} with radii within $1\sigma$ of the measured radius of \aumicc, as presented in Fig.~\ref{fig:mass_radius}. The median and median absolute deviation of the mass of exoplanets in this population is $0.043\pm0.036$~\MJ. Therefore, the mass of \aumicc\ is likely to lie in the range of 0.007~\MJ$<M_{\rm c}<$0.079~\MJ, implying a planet density in the range of 0.4~g\,cm$^{-3}<\rho_{\rm c}<$4.1~g\,cm$^{-3}$. Several different scenarios are possible for the internal structure of this planet depending on its mass. We estimate the semi-amplitude of the induced radial velocity caused by \aumicc\ in the star motion to be in the range of 0.8~m\,s$^{-1}<K_{\rm c}<$9.5~m\,s$^{-1}$. A number of world-class spectrometers can currently achieve the necessary precision to detect such reflex motion. However, the intense stellar activity in \aumic\ may represent a challenge for such measurement.

In order to analyze the stability of the orbital solution (Table~\ref{tab:derivedparams}), we performed a global frequency analysis \citep{Laskar_1990, Laskar_1993PD} in the vicinity of the best fit, in the same way as achieved for other planetary systems \citep[e.g.,][]{Correia_etal_2005, Correia_etal_2010}. The system is integrated on a regular 2D mesh of initial conditions, with varying semi-major axis and eccentricity of planet~$c$, while the other parameters are retained at their nominal values. We used the symplectic integrator SABA1064 of \citet{Farres_etal_2013}, with a step size of $5\times 10^{-3}$~yr  and general relativity corrections. Each initial condition is integrated over 5~kyr , and a stability indicator is computed to be the variation in the measured mean motion over the two consecutive 2.5~kyr intervals of time \citep[for more details see][]{Couetdic_etal_2010}. For regular motion, there is no significant variation in the mean motion along the trajectory, while it can vary significantly for chaotic trajectories.

\begin{figure}
   \centering
   \includegraphics[width=0.95\hsize]{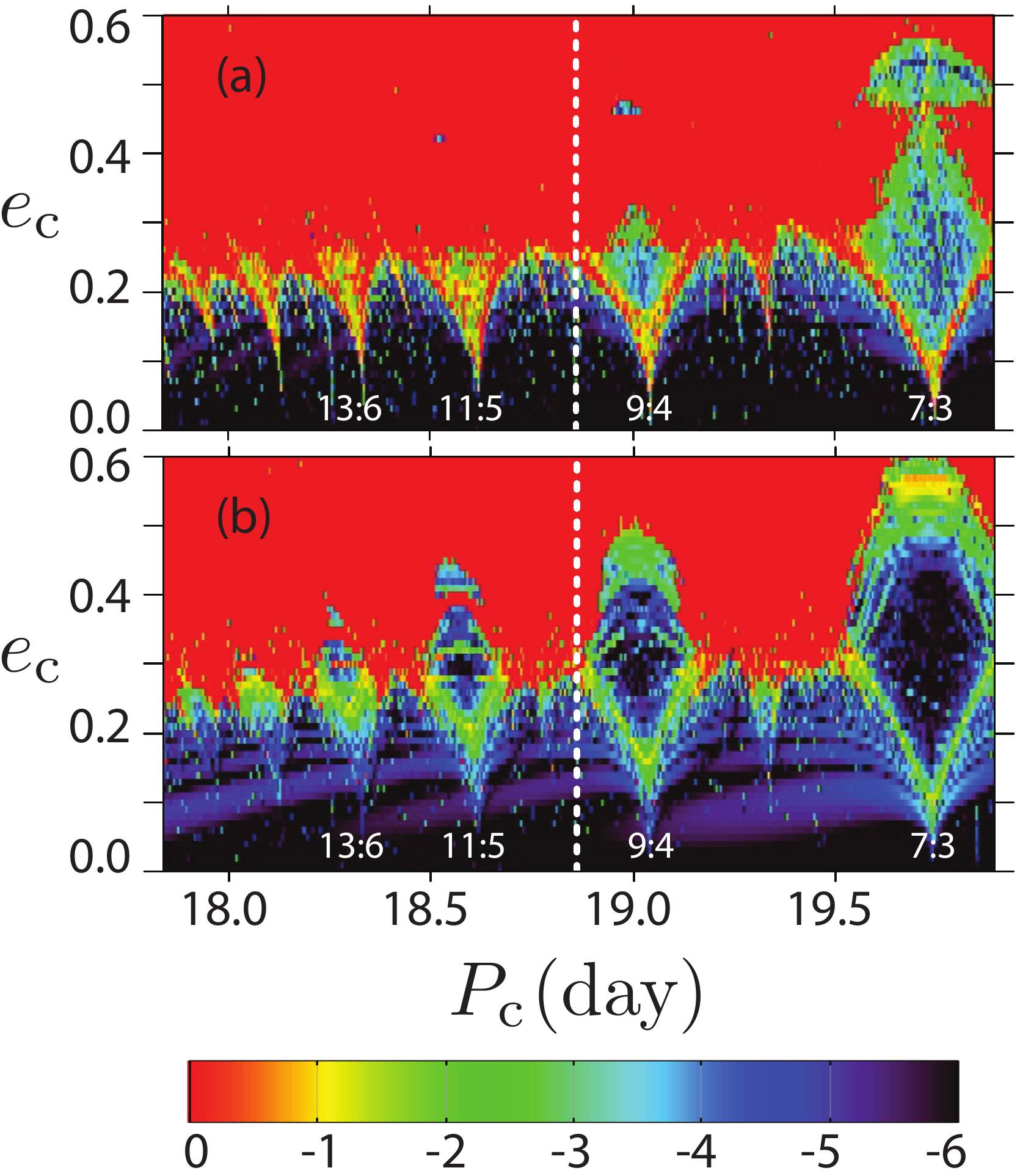}
      \caption{Stability analysis of the AU Mic planetary system, around the best fit solution (Table~\ref{tab:derivedparams}), with $M_{\rm c} = 25.0$~M$_\oplus$ (a), and $M_{\rm c} = 2.2$~M$_\oplus$ (b). The phase space of the system is explored by varying the semi-major axis $a_c$ (and thus the period $P_c$) and eccentricity $e_c$ of the outer planet; $e_c$ is plotted as a function of $P_c$ and not $a_c$, as the uncertainty on $a_c$ is larger.  For the initial conditions, the step size is $0.01$ in eccentricity, and $2\times 10^{-5}$~au  in semi-major axis. For each initial condition, the system is integrated over 5 kyr and a stability criterion is derived with the frequency analysis of the mean longitude. The chaotic diffusion is measured by the variation in the mean motion. The color scale corresponds to the decimal logarithm of the variation of the mean motion \citep{Correia_etal_2010}. The ``red'' zone corresponds to highly unstable orbits, while  the ``dark-blue'' region can be assumed to be stable on a billion-years timescale. The main resonances in the vicinity of the solution (13:6, 11:5, 9:4, 7:3) are labeled. It should be noted that for the lower value of the mass $M_{\rm c}$ (b), the libration island is mostly stable while it is not the case for the larger value of $M_{\rm c}$ (a).}
        \label{fig:stab}
  \end{figure}

In Fig.~\ref{fig:stab}, we show the wide vicinity of the nominal solution for two different configurations, one with $M_{\rm c} = 25.0$~M$_\oplus$ (top) and another with $M_{\rm c} = 2.2$~M$_\oplus$ (bottom), corresponding to the maximum and minimum masses of the outer planet, respectively.  The stability indicator is reported using a color index, where ``red'' represents the strongly chaotic trajectories and ``dark-blue''  shows the extremely stable ones.  We observe that  there are several islands of mean motion resonances  nearby, however, the main difference is that when the mass of the outer planet is large, these resonances become unstable.  The best fit nominal solution (vertical dotted line) is close to the 9:4 resonance, but outside, and so the nominal solution is in the stable zone for both mass values, as long as $e_c < 0.2$. We hence conclude that the AU Mic planetary system is stable and nearly circular.

The proximity to the 9:4 resonance could cause significant transit-timing variation (TTV).  To measure the TTV for both planets, we run an MCMC analysis for each individual transit observed by TESS, where we obtained an independent measurement of the times of transits. The transit parameters are fixed to their best fit values, except $T_{c}$, which is set with an uniform prior distribution of $\mathcal{U}(T_{c}-0.005,T_{c}+0.005)$, for $T_{c}$ being the time of conjunction calculated from the fit ephemeris. We subtract $T_{c}$ from the measured times of conjunction to obtain the TTVs as presented in Table \ref{tab:ttvs}. We detect no significant TTV in the TESS data. TTVs with amplitudes of one minute or more would have been likely detected in this dataset.

\begin{table}
\centering
\caption{Measurements of the transit-timing variation (TTV) for the transits of \aumicb\ and c observed by TESS.}
\label{tab:ttvs}
\begin{tabular}{cccc}
\hline
Planet & epoch & $T_{c}$ (TBJD) & TTV (s)\\
\hline
b & 0 & $1330.39046\pm0.00016$ & $-32\pm17$\\
b & 2 & $1347.31646\pm0.00016$ & $+39\pm21$\\
b & 84 & $2041.28238\pm0.00026$ & $-41\pm25$\\
b & 85 & $2049.74538\pm0.00026$ & $+31\pm32$\\
b & 86 & $2058.20838\pm0.00026$ & $+27\pm27$\\
\hline
c & 0 & $1342.22432\pm0.00031$ & $-13\pm43$\\
c & 37 & $2040.00697\pm0.00048$ & $-81\pm53$\\
c & 38 & $2058.86596\pm0.00049$ & $+92\pm59$\\
\hline
\end{tabular}
\end{table}

\begin{figure}
   \centering
   \includegraphics[width=0.9\hsize]{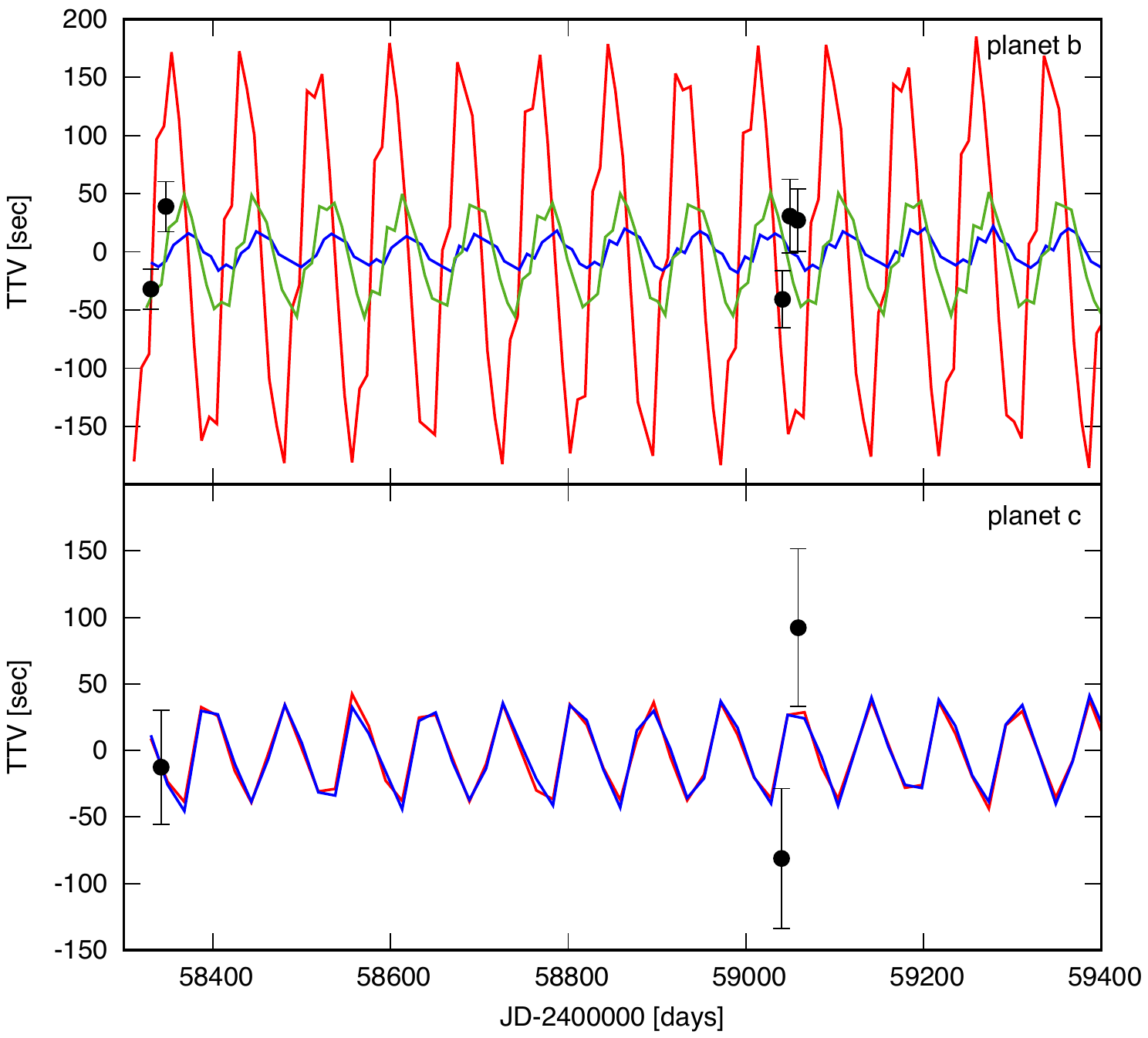}
      \caption{ Transit timing variations for \aumicb\ (top) and \aumicc\ (bottom). The colored lines correspond to synthetic data obtained with the best fit solution (Table~\ref{tab:derivedparams}) using different values for the mass of the outer planet: $M_{\rm c} = 2.2$~M$_\oplus$ (blue), $M_{\rm c} = 7.0$~M$_\oplus$ (green), and $M_{\rm c} = 25.0$~M$_\oplus$ (red). The dots correspond to the observed TTVs. (Table~\ref{tab:ttvs}).}
        \label{fig:ttv}
  \end{figure}

Using the nominal solution from  Table~\ref{tab:derivedparams}, we generated the synthetic TTVs as in \cite{hebrard2020}, for the minimum and the maximum masses of the outer planet.  In Fig.~\ref{fig:ttv}, we show the variations corresponding to each solution superimposed with the observational data (Table~\ref{tab:ttvs}). For the outer planet, the amplitude of the TTV is around one minute for both mass choices. However, for the inner planet, the minimum mass produces TTV with an amplitude of only a few seconds, while for the maximum mass the amplitude can reach up to three minutes. Thus, the TTV of the inner planet can place constraints on the mass of the outer planet.

The precision and the number of photometric measurements currently available for the \aumic\ system (Table~\ref{tab:ttvs}) do not allow us to run an exhaustive search for a best fit solution, but they allow us to reduce the uncertainty in the mass of the outer planet. The non-detection of TTVs that are larger than one minute place the mass of the outer planet on the low side as the likelier assumption.  In Fig.~\ref{fig:ttv}, we show the TTV corresponding to an outer companion with $M_{\rm c} = 7.0$~M$_\oplus$.  This mass generates TTVs with an amplitude around one minute corresponding to our upper limit. The observation of additional transits should help to resolve the present ambiguity in the outer planet mass.

Furthermore, we calculated the habitable zone (HZ) for \aumic\ using the equations and data from \cite{Kopparapu2014}, which gives an optimistic lower limit (recent Venus) at 0.25~au, and an upper limit (early Mars) at 0.64~au, with the runaway greenhouse limits ($M_{\rm p}=1$~\ME) ranging between 0.32~au and 0.61~au. \aumicb\ and c being at an orbital distance of $0.0645\pm0.0013$~au and $0.1101\pm0.0022$~au are not in the HZ. We estimate the equilibrium temperatures as in \cite{heng2013} assuming an uniform heat redistribution and an arbitrary geometric albedo of 0.1, which gives $T_{\rm eq, b}=593\pm21$~K and $T_{\rm eq, c}=454\pm16$~K, for \aumicb\ and c, respectively, showing that these warm Neptunes are indeed too hot to sustain water in liquid state. Moreover, the actual temperature for these planets could be much higher due to a possible greenhouse effect depending on their atmospheric composition and also due to an elevated internal temperature provided the young age of the system.


\section{Conclusions}
\label{sec:conclusions}

We presented an analysis of the photometric TESS observations of the active young M1 star \aumic, where we model the rotational modulation by starspots, the flaring activity, and the planetary transits simultaneously. Our analysis delivered an estimation of the flare occurrence rate of 6.35 flares per day for flares with amplitudes in the range of $0.06\% < f_{\rm max} < 1.5\%$ of the star flux. With such a flare rate it is important to model flaring activity to improve the constraints on the planetary parameters from transits. In our simple multi-flare model, we did not explore a detailed physical modeling of flares  \citep[e.g., as in][]{tilley2019} that could, in principle, be studied more extensively given the high quality and broad time coverage of these TESS observations. Our aim was to find an empirical description of flares that improves the constraints on the planetary parameters of \aumicb\ and that increases the sensitivity for detecting and characterizing the second planet \aumicc.

Our analysis of the five transits of \aumicb\ provided measurements of the orbital period and time of conjunction, giving an improved ephemeris of $T_{c} = 2\,458\,330.39051\pm0.00015 + E\times 8.463000\pm0.000002$~JD for an accurate prediction of the times of future transits.
Our detection of two new transits in the recent TESS observations have provided substantial evidence to confirm the second planet, that is, \aumicc. We analyzed the three transits and obtained consistent results for the planetary parameters, which supports our hypothesis that these transits come from the same planet.  Our MCMC analysis provided a determination of the planetary parameters for \aumicc, delivering an ephemeris of $T_{c} = 2\,458\,342.2223\pm0.0005 + E\times 18.859019^{+0.000016}_{-0.000015}$~JD for the times of transits with a transit duration of $4.5\pm0.8$ hours. The derived radius of $3.24\pm0.16$~\RE\ indicates that \aumicc\ is a Neptune-size planet with an expected mass in the range of 2.2~\ME$< M_{\rm c} <$25.0~\ME, estimated from the population of exoplanets of a similar size.

\aumicb\ and c with an orbital period of $8.46$~d and $18.86$~d are near the 9:4 mean-motion resonance. Such a configuration is stable and could cause significant TTV, which we do not detect with the current dataset. Still, that non-detection provides an upper limit to the mass of \aumicc\ of $M_{\rm c} < 7$~\ME, implying an upper limit to the planet density of $\rho_{\rm c}<$1.3~g\,cm$^{-3}$. This suggests that both planets are likely inflated, as it should be expected for young planets \citep[e.g.,][]{Helled2020}.  Therefore, these planets are interesting targets for atmospheric characterization by transmission spectroscopy.

The interactions between the planets coupled in a near 9:4 mean-motion resonance and the debris disk surrounding the star are to be the subject of subsequent studies. High-resolution images have revealed that the disk has a complex structure with small-sized substructures, such as "feature A" at a dozen astronomical units showing a ``loop-like'' morphology \citep{Wisniewski2019}. As for the case of $\beta$\,Pictoris, it has been shown that the planets can sculpt the disk morphology through the gravitational interaction between planets and the debris disk parents bodies at far distances \citep{Lecavelier1996,Augereau2001}, particularly through resonance mechanisms in young systems that are still evolving \citep{Lecavelier1998}. The impact of the presence of \aumicb\ and \aumicc\ on the dust disk deserves further analysis.

\begin{acknowledgements}
     We acknowledge funding from the French National Research Agency (ANR) under contract number ANR-18-CE31-0019 (SPlaSH). This paper includes data collected with the TESS mission, obtained from the MAST data archive at the Space Telescope Science Institute (STScI). In particular, the TESS data secured in 2020 are part of the TESS Guest Investigator Programs G03263 (PI: P. Plavchan), G03141 (PI: E. Newton), and G03273 (PI: L. Vega). Together with other data and analyses of that particularly interesting system, that will be the subjects of forthcoming papers by Cale et al., Collins et al., El Mufti et al., Gilbert et al., Wittrock et al. Funding for the TESS mission is provided by the NASA Explorer Program. STScI is operated by the Association of Universities for Research in Astronomy, Inc., under NASA contract NAS 5–26555. We thank the TESS Team members for making available the extremely accurate photometric data they obtained. We acknowledge support by CFisUC projects (UIDB/04564/2020 and UIDP/04564/2020), GRAVITY (PTDC/FIS-AST/7002/2020), ENGAGE SKA (POCI-01-0145-FEDER-022217), and PHOBOS (POCI-01-0145-FEDER-029932), funded by COMPETE 2020 and FCT, Portugal. This work benefited from HPC resources of MesoPSL financed by the Region Ile de France and the project Equip@Meso (reference ANR-10-EQPX-29-01) of the programme Investissements d’Avenir.

\end{acknowledgements}

%
%

\bibliographystyle{aa}

\bibliography{bibliography}

\clearpage
\onecolumn

\begin{appendix}

\section{Star rotation}
\label{app:starrotation}

In this appendix, we present an independent measurement of the star rotation period of \aumic.

\begin{figure}
   \centering
   \includegraphics[width=0.9\hsize]{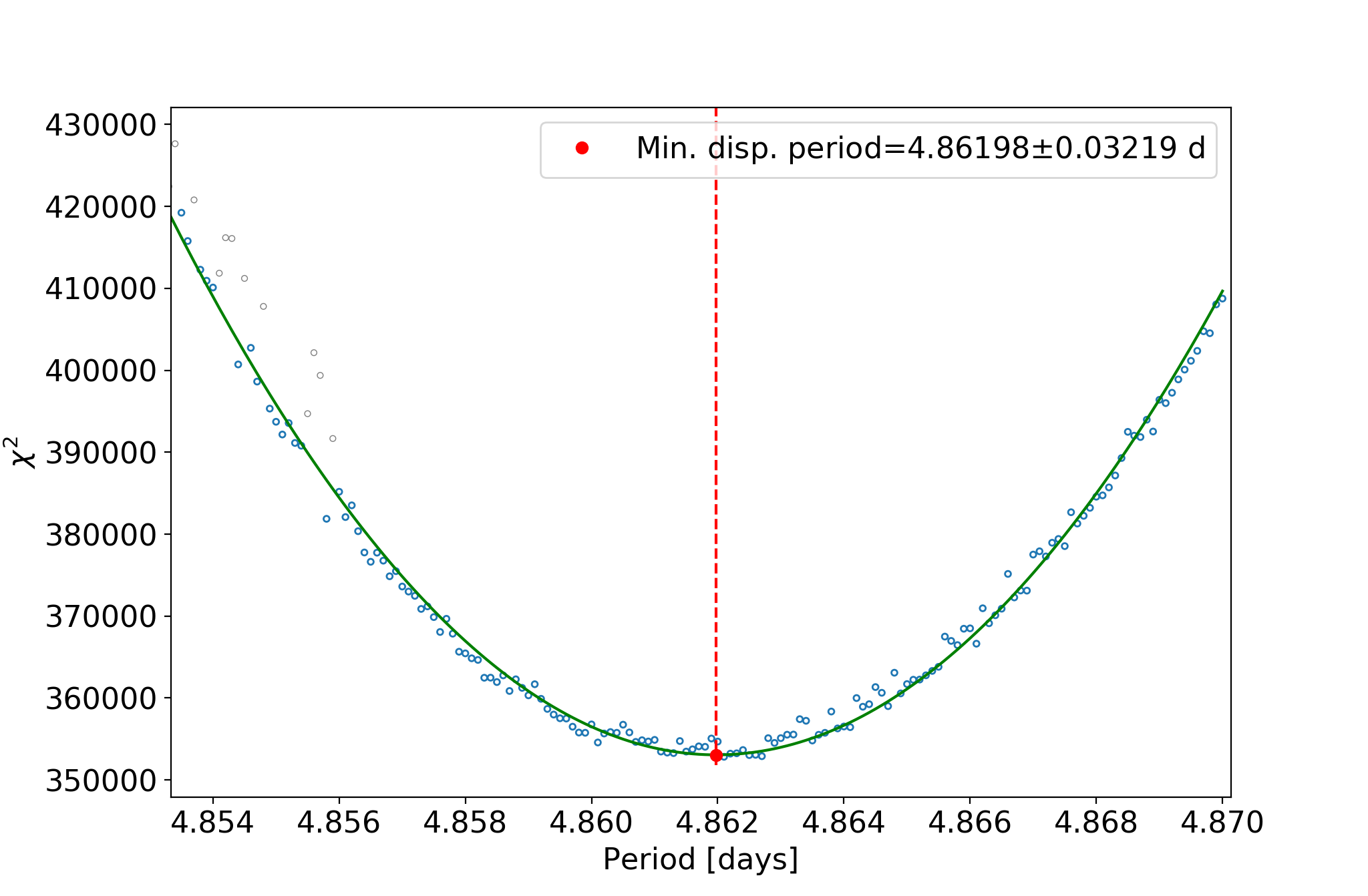}
      \caption{Rotation period versus chi-square for the reduced light curve of \aumic\, after removing flares and planetary transits. Red dashed line shows the minimum $\chi^2$ at $P=4.862\pm0.032$~d obtained by a parabolic fit (green line).
        }
        \label{fig:rotper}
  \end{figure}

\begin{figure}
   \centering
   \includegraphics[width=0.9\hsize]{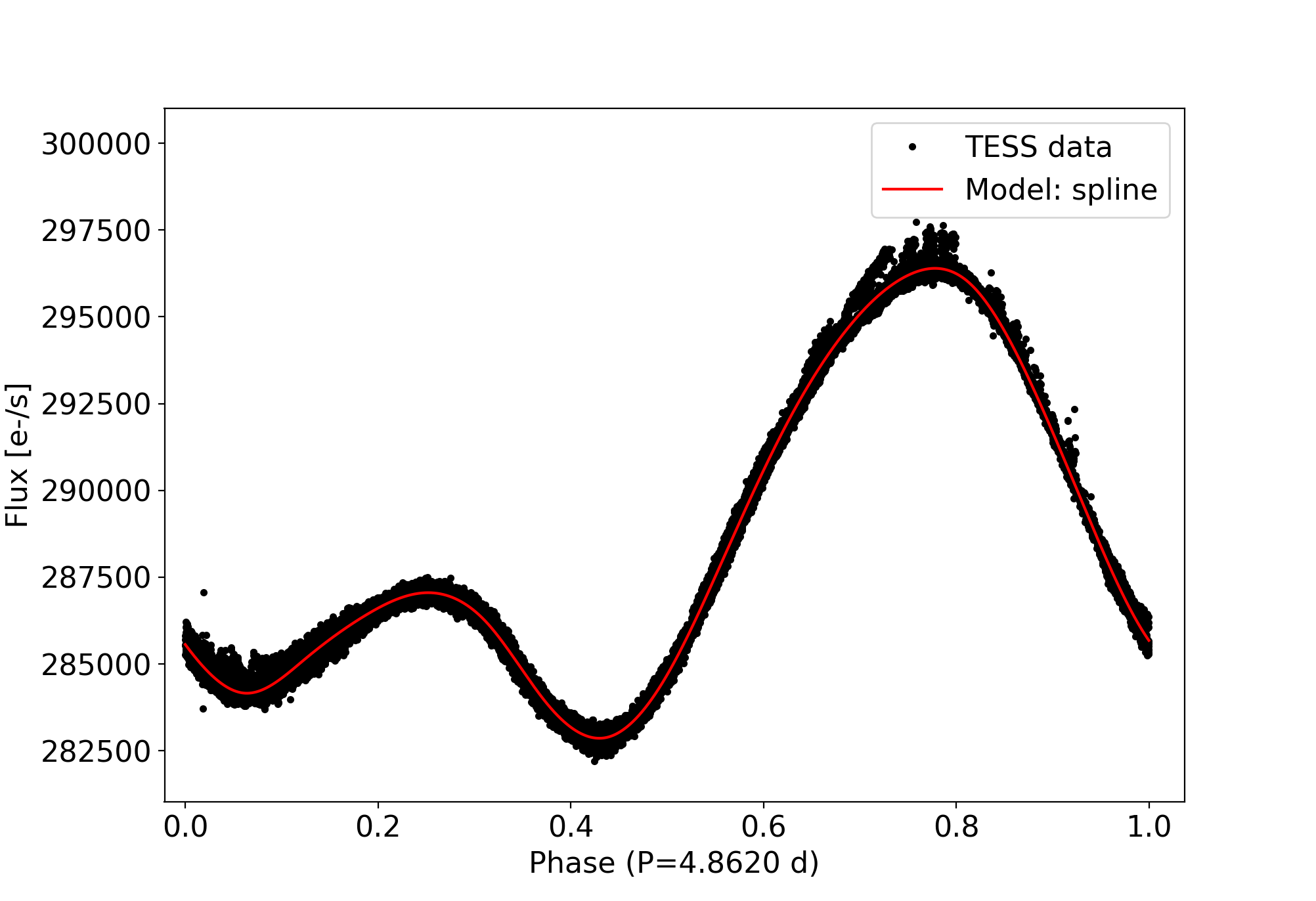}
      \caption{Phased light curve of \aumic\ using our measured rotation period of 4.862~d. Black points show the 2018 TESS (first visit only) fluxes after removing the flares and transits best fit models, and red line shows the best fit cubic spline model to the phased data.
        }
        \label{fig:aumicphasecurve}
  \end{figure}

Starspots in \aumic\ have a life time sufficiently large that the overall spot pattern did not change significantly during the 2018 TESS observations in the first visit. As one can see in the bottom panel of Fig.~\ref{fig:tesslc}, there has been a more significant change in the spot pattern during the second visit of TESS, where the amplitude of the cyclic flux variations decreases with time. For this reason, we did not include the second visit data to measure the rotation period.  Flares and planetary transits are removed from the original data using the models computed in Sects. \ref{sec:flares}, \ref{sec:aumicbtransits}, and \ref{sec:aumicc}. We apply the phase dispersion minimization (PDM) method inspired by the work of \cite{stellingwerf1978}, where we calculate the phase curve for several values of trial rotation periods ranging from 4.85~d to 4.88~d in steps of 0.0001~d. The rotation period of $P_{r}=4.865$~d measured by \cite{torres1972} was considered as initial guess to define our search range.  For each trial period, we fit a cubic spline with 15 knots and calculate the chi-square by $\chi^2=\sum{(F_{i}-F_c)^2/\sigma_{i}^2}$, for $F_{i}$ being the flux after removing flares and transits, $F_c$ being the fit model, and $\sigma_{i}$ being the flux uncertainty given by the TESS pipeline. The results are presented in Fig.~\ref{fig:rotper}, where we find a minimum $\chi^2$ at $P_r=4.862\pm0.032$~d, which is consistent but less accurate than the rotation period of $4.863\pm0.010$~d obtained by \cite{plavchan2020Natur} from the same data set. The resulting phase curve and best fit model are presented in Fig.~\ref{fig:aumicphasecurve}. The final dispersion of residuals is 265~e$^{-}$s$^{-1}$, which is more than twice the dispersion of residuals from the fit to the time series, that is, 99~e$^{-}$s$^{-1}$ for the data from the first visit of TESS only. This discrepancy is likely due to the evolution or drifts of starspots during the several rotation cycles covered in the first visit and perhaps also due to differential rotation \citep{klein2020}, which could generate a dispersion in the phase diagram since starspots at different latitudes are modulated by different periods.

Assuming the radius of \aumic\ measured by interferometry of $R_{\star}=0.75\pm0.03$~\RS\ \citep{russel2015} and the rotation period of $P_{r}=4.862\pm0.032$~d, we calculate the velocity at the equator as $v_{\rm eq} = 2 \pi R_{\star} / P_{r} = 7.8\pm0.3$~km\,s$^{-1}$. This gives a maximum value of the projected velocity of $v_{\rm eq}\sin{(i_{\star})}<8.1$~km\,s$^{-1}$, for $i_{\star}$ being the inclination angle of the rotation axis of the star with respect to the line of sight.

\section{Flares fit parameters}
\label{app:flaresfitparams}

\begin{longtable}{rcrr}
\caption{Flares fit parameters.}\\
\label{tab:flaresfitparams}
Flare  & t$_{\rm peak}$ & amplitude & FWHM \\
index  & TBJD            & e$^{-}$s$^{-1}$ & min    \\
\hline
1 & 1325.5892 & 427.5 & 31.84 \\
2 & 1325.7407 & 413.9 & 2.12 \\
3 & 1326.1596 & 284.5 & 16.06 \\
4 & 1326.2210 & 299.1 & 1.58 \\
5 & 1326.3398 & 167.0 & 44.57 \\
6 & 1326.3683 & 309.0 & 18.73 \\
7 & 1326.5566 & 235.3 & 6.42 \\
8 & 1326.6753 & 2121.5 & 20.65 \\
9 & 1326.6978 & 2553.5 & 41.60 \\
10 & 1326.7152 & 643.7 & 30.53 \\
11 & 1326.8131 & 407.7 & 17.67 \\
12 & 1327.1280 & 182.7 & 7.49 \\
13 & 1327.2165 & 710.9 & 40.91 \\
14 & 1327.3085 & 234.8 & 11.85 \\
15 & 1327.4389 & 475.8 & 11.59 \\
16 & 1327.4702 & 618.6 & 6.97 \\
17 & 1327.5782 & 154.6 & 81.53 \\
18 & 1327.8975 & 1430.4 & 8.48 \\
19 & 1327.9450 & 405.8 & 100.61 \\
20 & 1327.9970 & 259.6 & 35.31 \\
21 & 1328.1461 & 537.6 & 10.28 \\
22 & 1328.2454 & 323.7 & 7.80 \\
23 & 1328.5056 & 7272.7 & 7.27 \\
24 & 1328.7362 & 1838.0 & 3.24 \\
25 & 1329.0420 & 301.9 & 17.15 \\
26 & 1329.0900 & 378.9 & 14.88 \\
27 & 1329.3240 & 689.1 & 13.58 \\
28 & 1329.3803 & 2870.0 & 7.23 \\
29 & 1329.4333 & 861.4 & 13.85 \\
30 & 1329.4581 & 168.4 & 23.54 \\
31 & 1330.0446 & 221.5 & 10.54 \\
32 & 1330.1271 & 1112.7 & 25.56 \\
33 & 1330.3750 & 338.9 & 36.00 \\
34 & 1330.4089 & 295.2 & 5.27 \\
35 & 1330.4475 & 184.4 & 71.83 \\
36 & 1330.4524 & 1517.1 & 4.97 \\
37 & 1330.6290 & 308.9 & 41.26 \\
38 & 1330.8089 & 793.3 & 37.58 \\
39 & 1330.9154 & 542.9 & 41.27 \\
40 & 1331.3989 & 576.3 & 37.30 \\
41 & 1331.4522 & 383.9 & 45.27 \\
42 & 1331.8229 & 1486.5 & 3.80 \\
43 & 1332.0069 & 178.2 & 20.97 \\
44 & 1332.1019 & 195.6 & 23.99 \\
45 & 1332.3460 & 536.7 & 9.86 \\
46 & 1332.4868 & 371.5 & 105.57 \\
47 & 1332.8700 & 299.1 & 12.73 \\
48 & 1333.0207 & 504.1 & 6.85 \\
49 & 1333.1734 & 617.6 & 22.72 \\
50 & 1333.7962 & 288.6 & 36.04 \\
51 & 1333.8896 & 5881.7 & 17.38 \\
52 & 1333.8961 & 1535.9 & 9.14 \\
53 & 1333.9019 & 1693.0 & 1.99 \\
54 & 1334.1192 & 2089.1 & 13.44 \\
55 & 1334.3769 & 811.2 & 5.82 \\
56 & 1334.3925 & 442.6 & 5.83 \\
57 & 1334.4750 & 172.1 & 61.01 \\
58 & 1334.6162 & 527.2 & 9.53 \\
59 & 1334.6378 & 223.0 & 8.21 \\
60 & 1335.1163 & 11469.8 & 13.44 \\
61 & 1335.1293 & 3267.2 & 11.25 \\
62 & 1335.1639 & 1568.5 & 119.59 \\
63 & 1335.3617 & 12798.6 & 5.00 \\
64 & 1335.7034 & 3266.6 & 4.68 \\
65 & 1335.7127 & 2214.6 & 0.65 \\
66 & 1335.7684 & 206.0 & 16.34 \\
67 & 1335.8508 & 202.2 & 17.09 \\
68 & 1336.1992 & 434.9 & 20.97 \\
69 & 1336.2313 & 1054.6 & 6.61 \\
70 & 1336.2723 & 246.9 & 6.35 \\
71 & 1336.3060 & 330.1 & 6.64 \\
72 & 1336.5145 & 506.0 & 4.03 \\
73 & 1336.5422 & 447.8 & 25.73 \\
74 & 1336.6280 & 384.5 & 6.94 \\
75 & 1336.9576 & 2572.0 & 9.42 \\
76 & 1337.0216 & 1119.7 & 9.59 \\
77 & 1337.0341 & 2277.9 & 21.60 \\
78 & 1337.4779 & 681.6 & 8.05 \\
79 & 1337.6259 & 473.3 & 3.50 \\
80 & 1337.7419 & 355.1 & 16.21 \\
81 & 1337.9638 & 346.6 & 18.35 \\
82 & 1338.0760 & 215.2 & 4.12 \\
83 & 1338.1343 & 190.1 & 7.34 \\
84 & 1338.2252 & 1050.7 & 11.79 \\
85 & 1338.2467 & 480.7 & 79.62 \\
86 & 1338.4465 & 1526.2 & 2.42 \\
87 & 1338.5024 & 1309.5 & 6.00 \\
88 & 1338.5213 & 2095.5 & 18.37 \\
89 & 1339.7341 & 200.4 & 9.29 \\
90 & 1339.7816 & 466.4 & 34.29 \\
91 & 1339.8383 & 1029.1 & 10.20 \\
92 & 1340.5251 & 831.2 & 37.34 \\
93 & 1340.6133 & 399.3 & 26.63 \\
94 & 1340.9817 & 262.2 & 12.08 \\
95 & 1341.3729 & 390.8 & 34.20 \\
96 & 1341.9796 & 651.3 & 4.16 \\
97 & 1342.1527 & 234.9 & 33.97 \\
98 & 1342.1668 & 1938.2 & 0.72 \\
99 & 1342.2354 & 224.3 & 12.48 \\
100 & 1342.2719 & 209.6 & 18.33 \\
101 & 1342.2962 & 365.4 & 1.86 \\
102 & 1342.4018 & 122.2 & 39.11 \\
103 & 1342.6043 & 421.3 & 4.31 \\
104 & 1342.6336 & 255.9 & 24.70 \\
105 & 1343.5102 & 1152.4 & 10.70 \\
106 & 1343.9443 & 221.4 & 67.41 \\
107 & 1344.0673 & 308.4 & 8.81 \\
108 & 1344.2179 & 185.1 & 18.45 \\
109 & 1344.5174 & 1246.6 & 3.87 \\
110 & 1344.5350 & 699.4 & 11.15 \\
111 & 1344.7107 & 272.2 & 88.42 \\
112 & 1344.8182 & 782.1 & 18.63 \\
113 & 1345.0144 & 474.4 & 14.07 \\
114 & 1345.0717 & 405.8 & 3.72 \\
115 & 1345.2198 & 425.1 & 23.96 \\
116 & 1345.2599 & 296.4 & 23.24 \\
117 & 1345.3944 & 2205.4 & 6.26 \\
118 & 1345.4257 & 1232.8 & 24.61 \\
119 & 1345.4816 & 329.7 & 24.13 \\
120 & 1345.5112 & 1299.7 & 19.20 \\
121 & 1345.5269 & 1908.0 & 31.64 \\
122 & 1345.5806 & 512.1 & 39.17 \\
123 & 1345.8324 & 526.4 & 3.90 \\
124 & 1346.0371 & 756.7 & 11.13 \\
125 & 1346.1915 & 1815.8 & 11.52 \\
126 & 1346.2062 & 622.6 & 42.72 \\
127 & 1346.2819 & 404.2 & 25.73 \\
128 & 1346.4695 & 1355.0 & 0.72 \\
129 & 1346.5211 & 5102.5 & 4.92 \\
130 & 1346.5619 & 1014.5 & 9.39 \\
131 & 1346.8081 & 311.1 & 21.61 \\
132 & 1346.8791 & 221.4 & 2.58 \\
133 & 1347.1743 & 224.4 & 49.74 \\
134 & 1347.2600 & 436.4 & 7.55 \\
135 & 1347.2696 & 773.2 & 7.27 \\
136 & 1347.2753 & 319.5 & 8.28 \\
137 & 1347.3391 & 491.7 & 2.12 \\
138 & 1347.4334 & 396.3 & 9.55 \\
139 & 1347.5908 & 326.9 & 19.73 \\
140 & 1348.4951 & 445.8 & 42.47 \\
141 & 1348.5686 & 479.9 & 7.37 \\
142 & 1348.5951 & 55.8 & 0.01 \\
143 & 1349.4535 & 1209.8 & 8.58 \\
144 & 1349.8225 & 855.3 & 24.96 \\
145 & 1350.0212 & 2444.4 & 23.85 \\
146 & 1350.0812 & 1351.2 & 176.66 \\
147 & 1350.1463 & 5308.9 & 4.98 \\
148 & 1350.2853 & 1769.8 & 11.17 \\
149 & 1350.2957 & 2460.0 & 12.99 \\
150 & 1350.3062 & 588.5 & 10.67 \\
151 & 1350.3212 & 392.3 & 7.01 \\
152 & 1350.5975 & 595.6 & 18.42 \\
153 & 1350.7887 & 347.2 & 4.36 \\
154 & 1351.7021 & 258.3 & 49.28 \\
155 & 1352.0038 & 254.4 & 59.49 \\
156 & 1352.3524 & 1573.3 & 3.12 \\
157 & 1352.6254 & 191.1 & 31.84 \\
158 & 1352.7400 & 228.6 & 28.74 \\
159 & 1352.9386 & 1629.7 & 0.43 \\
160 & 1352.9528 & 486.5 & 18.27 \\
161 & 1352.9878 & 437.3 & 49.39 \\
162 & 1353.0282 & 250.0 & 13.94 \\
163 & 2036.7060 & 1678.0 & 2.81 \\
164 & 2036.9803 & 557.7 & 22.15 \\
165 & 2037.0062 & 731.2 & 8.63 \\
166 & 2037.0445 & 528.1 & 119.19 \\
167 & 2037.4443 & 2482.4 & 5.59 \\
168 & 2037.4926 & 529.0 & 6.16 \\
169 & 2037.5567 & 629.3 & 0.58 \\
170 & 2037.6398 & 455.7 & 1.57 \\
171 & 2037.6661 & 285.1 & 0.07 \\
172 & 2037.6752 & 11108.5 & 2.09 \\
173 & 2037.6782 & 1725.8 & 10.30 \\
174 & 2037.6794 & 2650.0 & 22.55 \\
175 & 2037.7202 & 4475.7 & 3.33 \\
176 & 2037.7460 & 313.6 & 9.43 \\
177 & 2038.1988 & 160.4 & 44.15 \\
178 & 2039.1836 & 4086.6 & 4.74 \\
179 & 2039.2663 & 1637.8 & 1.89 \\
180 & 2039.3501 & 3249.4 & 5.30 \\
181 & 2039.3868 & 298.5 & 8.35 \\
182 & 2039.4411 & 733.2 & 10.44 \\
183 & 2039.5980 & 502.6 & 19.24 \\
184 & 2039.8267 & 522.3 & 17.25 \\
185 & 2039.8213 & 1015.2 & 3.54 \\
186 & 2039.9915 & 538.1 & 176.11 \\
187 & 2040.0245 & 560.2 & 20.59 \\
188 & 2040.3473 & 403.2 & 1.96 \\
189 & 2040.3548 & 313.3 & 2.82 \\
190 & 2040.7923 & 594.5 & 20.55 \\
191 & 2041.0997 & 231.1 & 53.67 \\
192 & 2041.3700 & 871.5 & 3.82 \\
193 & 2041.6072 & 409.8 & 9.48 \\
194 & 2041.7500 & 118.8 & 33.72 \\
195 & 2041.9375 & 333.8 & 45.71 \\
196 & 2041.9876 & 387.4 & 56.35 \\
197 & 2042.0502 & 407.3 & 19.53 \\
198 & 2042.4858 & 5650.2 & 4.36 \\
199 & 2042.5031 & 3220.7 & 8.76 \\
200 & 2042.6066 & 532.7 & 8.93 \\
201 & 2042.6998 & 522.8 & 1.64 \\
202 & 2043.3819 & 228.0 & 2.52 \\
203 & 2043.7737 & 266.4 & 48.63 \\
204 & 2044.0924 & 532.7 & 13.80 \\
205 & 2044.3488 & 540.4 & 3.00 \\
206 & 2044.5147 & 371.3 & 30.05 \\
207 & 2044.5761 & 600.4 & 34.88 \\
208 & 2044.7252 & 313.4 & 33.65 \\
209 & 2045.0332 & 163.6 & 8.18 \\
210 & 2045.3312 & 610.8 & 2.30 \\
211 & 2045.6075 & 242.5 & 24.26 \\
212 & 2045.7409 & 408.9 & 3.56 \\
213 & 2045.8442 & 351.2 & 34.24 \\
214 & 2046.3703 & 316.6 & 24.52 \\
215 & 2046.5418 & 284.7 & 95.80 \\
216 & 2046.6323 & 682.6 & 40.90 \\
217 & 2046.8676 & 806.2 & 2.98 \\
218 & 2047.0374 & 191.9 & 6.36 \\
219 & 2047.0975 & 270.8 & 8.22 \\
220 & 2047.2069 & 184.4 & 0.04 \\
221 & 2047.2155 & 3395.1 & 7.50 \\
222 & 2047.2688 & 340.0 & 38.58 \\
223 & 2047.3482 & 2740.9 & 9.52 \\
224 & 2047.4462 & 891.1 & 48.97 \\
225 & 2047.5051 & 1200.7 & 3.31 \\
226 & 2047.5481 & 539.1 & 2.23 \\
227 & 2047.6301 & 1138.6 & 9.53 \\
228 & 2047.8328 & 1166.7 & 1.96 \\
229 & 2048.0525 & 374.4 & 7.57 \\
230 & 2049.2135 & 206.6 & 26.42 \\
231 & 2049.5213 & 216.6 & 4.62 \\
232 & 2049.7238 & 206.7 & 3.53 \\
233 & 2049.7380 & 2365.4 & 2.16 \\
234 & 2049.7424 & 274.2 & 13.64 \\
235 & 2049.7546 & 468.5 & 8.35 \\
236 & 2049.9163 & 84.0 & 182.42 \\
237 & 2049.7972 & 280.8 & 4.41 \\
238 & 2050.3662 & 460.2 & 11.07 \\
239 & 2050.6606 & 453.6 & 18.94 \\
240 & 2050.7854 & 309.4 & 36.35 \\
241 & 2050.8613 & 1335.7 & 14.50 \\
242 & 2050.9074 & 396.0 & 22.97 \\
243 & 2051.1673 & 928.6 & 12.95 \\
244 & 2051.1775 & 523.8 & 18.06 \\
245 & 2051.4456 & 2285.7 & 10.79 \\
246 & 2051.4575 & 1210.5 & 40.84 \\
247 & 2051.5155 & 804.9 & 22.33 \\
248 & 2051.5398 & 491.4 & 66.83 \\
249 & 2051.8395 & 846.2 & 7.55 \\
250 & 2052.2513 & 387.8 & 5.20 \\
251 & 2052.3527 & 337.5 & 18.29 \\
252 & 2052.6859 & 526.4 & 28.58 \\
253 & 2052.8846 & 227.8 & 18.12 \\
254 & 2053.0858 & 588.8 & 2.19 \\
255 & 2053.2934 & 416.3 & 131.31 \\
256 & 2053.4293 & 11831.0 & 3.11 \\
257 & 2053.4313 & 4392.0 & 5.98 \\
258 & 2053.4364 & 2771.1 & 21.72 \\
259 & 2053.4471 & 1727.0 & 29.15 \\
260 & 2053.5172 & 655.4 & 27.49 \\
261 & 2053.7714 & 642.8 & 3.95 \\
262 & 2053.8511 & 541.5 & 7.36 \\
263 & 2054.2890 & 2125.9 & 40.06 \\
264 & 2054.5200 & 247.5 & 416.71 \\
265 & 2055.0243 & 356.7 & 25.01 \\
266 & 2055.1195 & 139.0 & 16.47 \\
267 & 2055.3076 & 305.4 & 3.46 \\
268 & 2055.4149 & 370.8 & 10.58 \\
269 & 2055.5277 & 365.8 & 9.81 \\
270 & 2055.6639 & 278.9 & 66.80 \\
271 & 2055.8219 & 257.3 & 34.42 \\
272 & 2055.9072 & 1427.7 & 23.05 \\
273 & 2055.9560 & 429.0 & 5.31 \\
274 & 2056.2248 & 219.2 & 39.11 \\
275 & 2056.2702 & 4218.7 & 1.21 \\
276 & 2056.3009 & 648.6 & 22.67 \\
277 & 2056.3696 & 432.8 & 39.37 \\
278 & 2056.4690 & 416.0 & 38.42 \\
279 & 2056.5167 & 515.0 & 15.28 \\
280 & 2056.6882 & 814.3 & 3.02 \\
281 & 2056.7114 & 411.5 & 3.34 \\
282 & 2056.8307 & 3072.5 & 7.83 \\
283 & 2056.8397 & 1642.1 & 23.47 \\
284 & 2056.8744 & 556.6 & 35.16 \\
285 & 2056.9135 & 5101.0 & 4.46 \\
286 & 2056.9178 & 6950.3 & 21.77 \\
287 & 2056.9413 & 1903.9 & 72.48 \\
288 & 2057.0256 & 1115.2 & 165.60 \\
289 & 2057.1330 & 855.9 & 7.08 \\
290 & 2057.2076 & 727.6 & 10.93 \\
291 & 2057.3293 & 335.9 & 1.80 \\
292 & 2057.6220 & 176.0 & 15.67 \\
293 & 2057.7999 & 8926.0 & 4.19 \\
294 & 2057.8206 & 805.8 & 35.86 \\
295 & 2058.0758 & 474.4 & 18.58 \\
296 & 2058.1089 & 341.1 & 2.95 \\
297 & 2058.2392 & 1637.1 & 10.08 \\
298 & 2058.2417 & 1628.0 & 2.74 \\
299 & 2058.2491 & 711.4 & 24.72 \\
300 & 2058.2591 & 642.2 & 20.10 \\
301 & 2058.2672 & 302.0 & 11.75 \\
302 & 2058.3522 & 260.7 & 21.24 \\
303 & 2058.7099 & 937.1 & 2.95 \\
304 & 2058.7918 & 258.1 & 15.96 \\
305 & 2058.8039 & 265.8 & 11.94 \\
306 & 2058.8144 & 272.5 & 15.28 \\
307 & 2058.8355 & 4984.7 & 0.16 \\
308 & 2058.8387 & 3972.5 & 7.65 \\
309 & 2058.8522 & 640.5 & 0.45 \\
310 & 2058.9145 & 564.9 & 0.32 \\
311 & 2058.9336 & 276.6 & 2.32 \\
312 & 2058.9386 & 365.7 & 1.12 \\
313 & 2059.0986 & 473.3 & 1.64 \\
314 & 2059.1116 & 628.6 & 58.56 \\
315 & 2059.3548 & 477.5 & 9.92 \\
316 & 2059.3865 & 520.4 & 9.98 \\
317 & 2059.4226 & 274.5 & 4.58 \\
318 & 2059.5478 & 4219.3 & 1.21 \\
319 & 2059.5512 & 3627.0 & 6.71 \\
320 & 2059.6466 & 351.4 & 16.89 \\
321 & 2059.6786 & 504.1 & 11.39 \\
322 & 2060.2895 & 493.0 & 1.21 \\
323 & 2060.4086 & 913.3 & 2.06 \\
324 & 2060.5060 & 851.8 & 24.80 \\
\hline
\end{longtable}

\section{MCMC samples and posterior distributions}
\label{app:posteriordistributions}

  \begin{figure*}
   \centering
   \includegraphics[width=0.9\hsize]{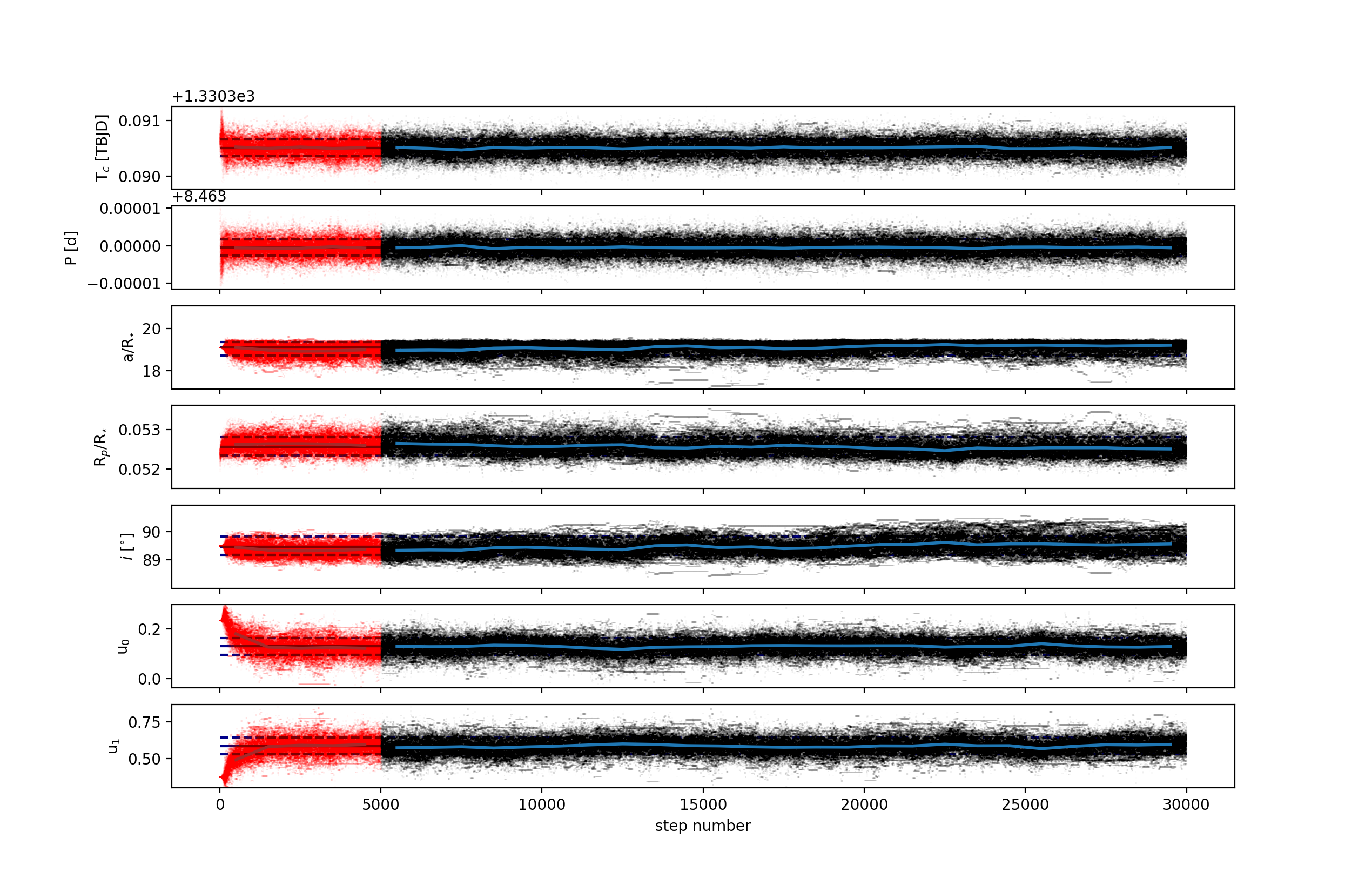}
      \caption{MCMC samples (black dots) used to calculate the posterior distributions of the transit parameters of \aumicb. The red dots show the discarded samples corresponding to the first 5000 steps of the chain. The blue lines show the median calculated for bins of 1000 steps. Dashed lines show the 16- and 84-percentiles and the solid horizontal lines shows the medians of the valid samples.)
         }
        \label{fig:aumic_b_mcmc_chain}
  \end{figure*}

  \begin{figure*}
   \centering
   \includegraphics[width=0.9\hsize]{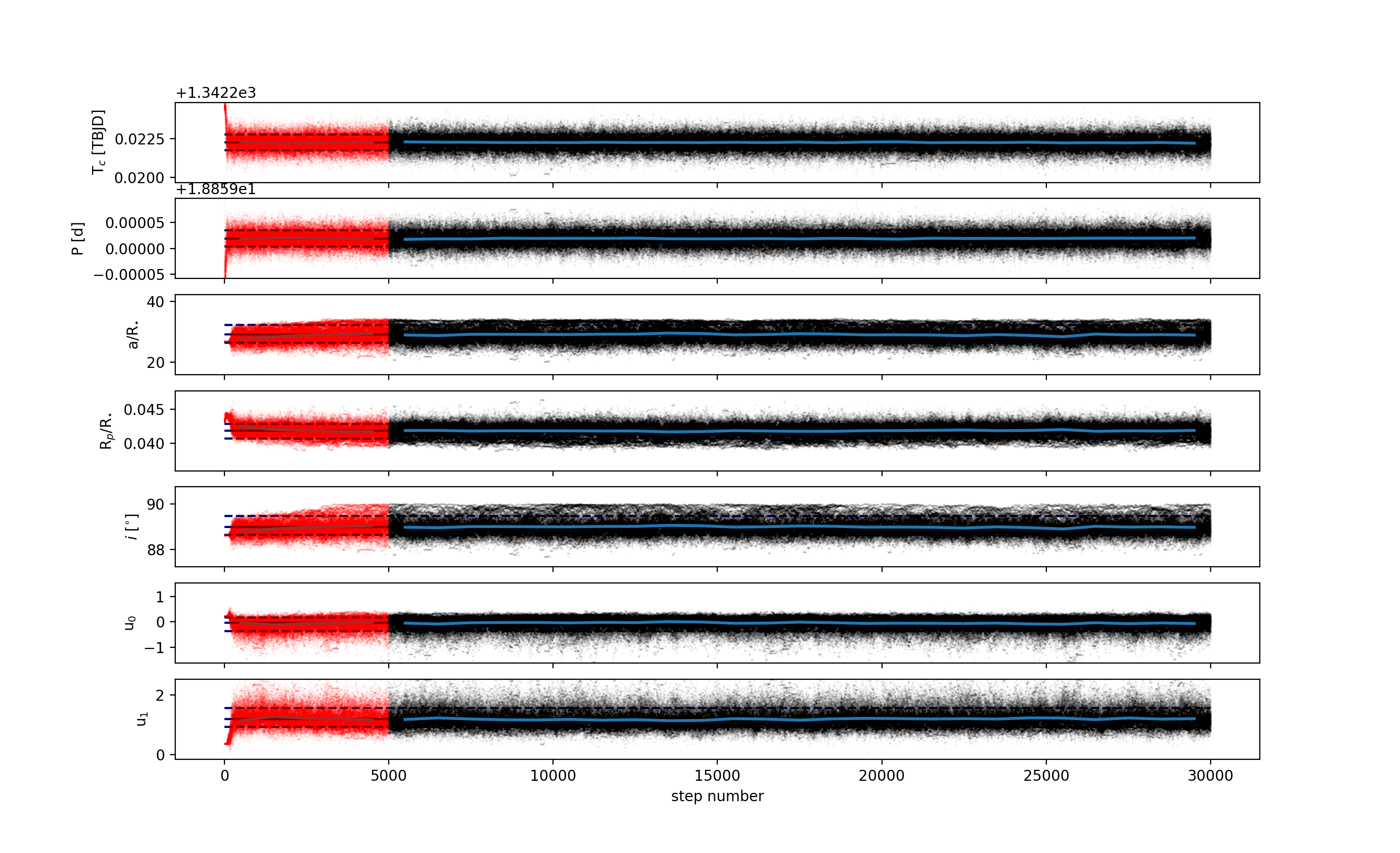}
      \caption{Same Fig.~\ref{fig:aumic_b_mcmc_chain} for the MCMC samples of the transit parameters of \aumicc.
         }
        \label{fig:aumic_c_mcmc_chain}
  \end{figure*}

  \begin{figure*}
   \centering
   \includegraphics[width=0.9\hsize]{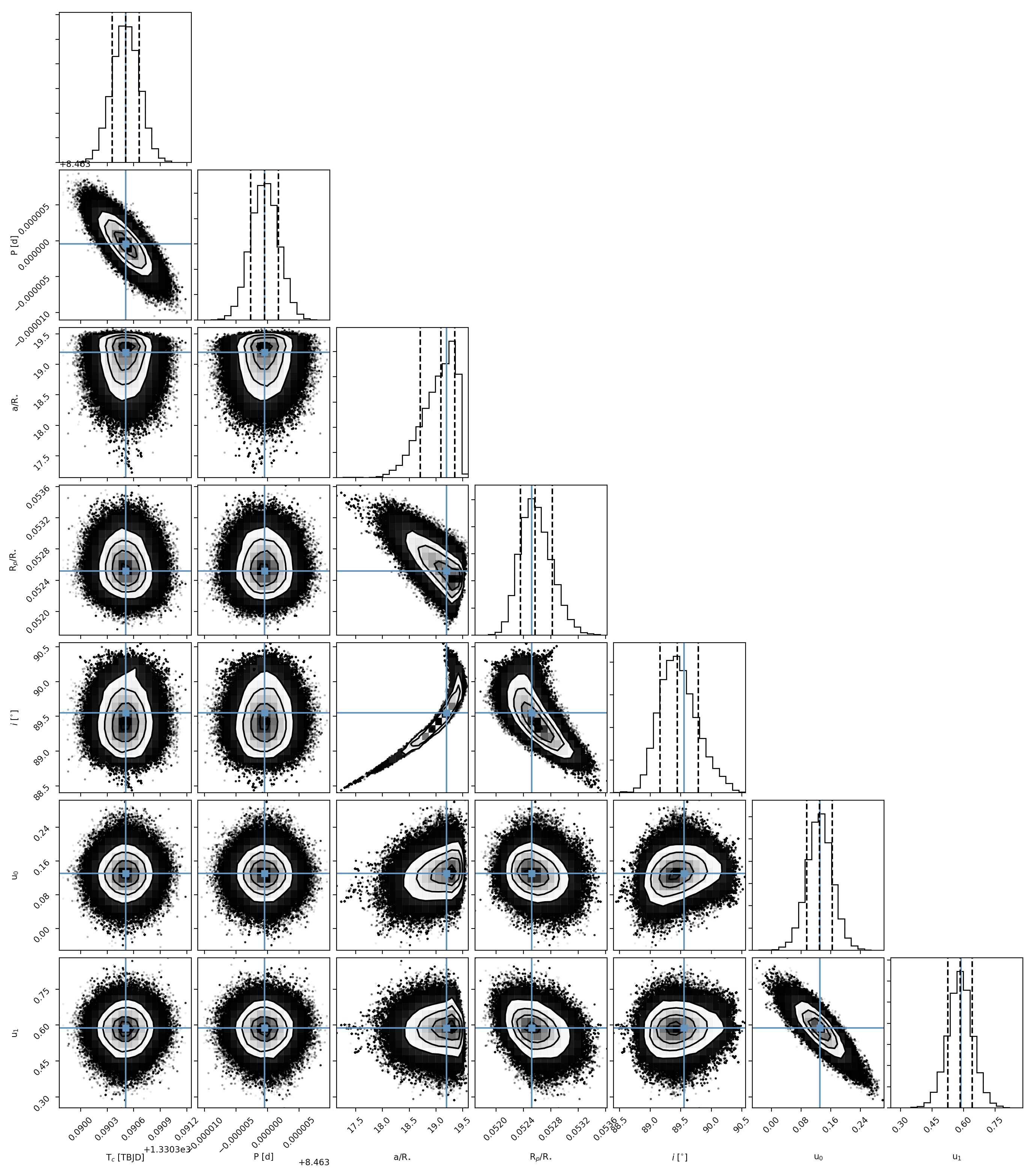}
      \caption{Pairs plot showing the MCMC samples and posterior distributions for the transit parameters of \aumicb\ as presented in Table \ref{tab:aumicbfitparams}. The contours mark the 1$\sigma$, 2$\sigma$, and 3$\sigma$ regions of the distribution. The blue crosses indicate the best fit values for each parameter and the dashed vertical lines in the projected distributions show the median values and the 1$\sigma$ uncertainty (34\% on each side of the median).
         }
        \label{fig:aumic_b_pairsplot}
  \end{figure*}

  \begin{figure*}
   \centering
   \includegraphics[width=0.9\hsize]{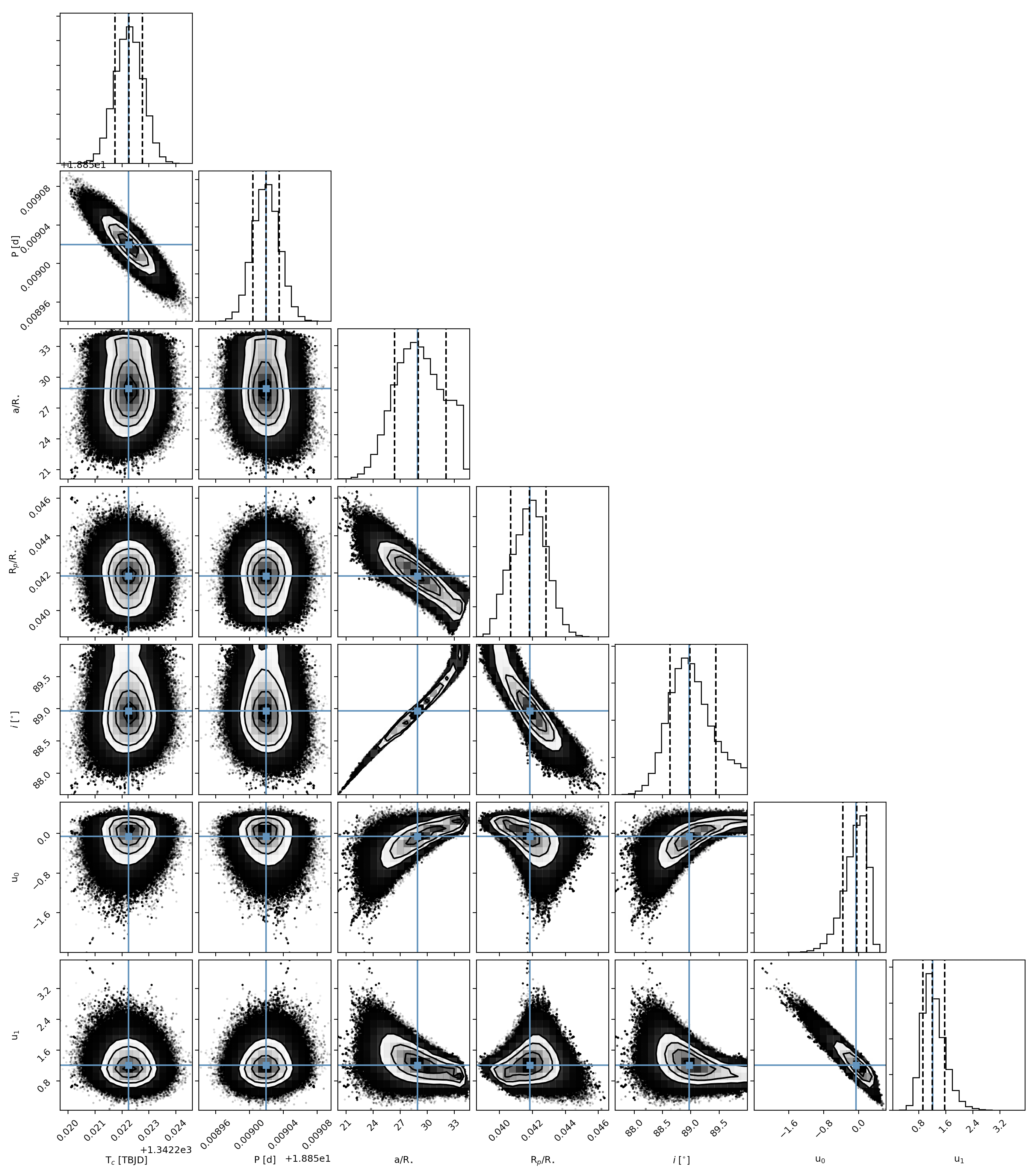}
      \caption{Pairs plot showing the MCMC samples and posterior distributions
for the transit parameters of \aumicb\ as presented in Table \ref{tab:aumicbfitparams}. Details are the same as in Fig.~\ref{fig:aumic_b_pairsplot} for the analysis of the three transits of \aumicc.
         }
        \label{fig:aumic_c_pairsplot}
  \end{figure*}

\section{Independent fit to the transits of \aumicc}
\label{app:individualtransitsofaumicc}

This appendix presents the results obtained from an MCMC analysis (as in Sect. \ref{sec:aumicbtransits}) performed on each transit of \aumicc\ independently. \ref{tab:othertransitsfitparams} shows the posterior of the transit parameters, where we present first the results using broad priors and then using the more restricted priors, as described in Sect. \ref{sec:aumicc}.  We notice that the period and semi-major axis are not well constrained by the shape of the transits. As a consequence, when using broad priors, their posteriors turn out to have relatively large uncertainties. However, the posterior of all parameters agree within $3\sigma$ among the three events and they also agree with our derived period of $P\sim18.86$-d.  The fit parameters for the more constrained priors also agree within $3\sigma$, and the dispersion of residuals are improved with respect to the unconstrained solution, which supports our hypothesis that these events are caused by the transits of the same planet, namely, \aumicc.

  \begin{table*}
    \caption[]{Fit transit parameters for an independent analysis of each of the three transit-like events identified as \aumicc\ in the TESS light curve. The first row shows the previous results for Event 1 by \cite{plavchan2020Natur}, the three following rows show the best fit parameters from our MCMC analysis of each event adopting an orbital period with an uniform prior of $P=\mathcal{U}(1,500)$~d. The last three rows show the fit parameters using an orbital period with a normal prior of $P=\mathcal{N}(18.85895,0.00003)$~d, as described in the text.}
    \label{tab:othertransitsfitparams}
    \begin{center}
    \begin{tabular}{cccccccc}
        \hline
        \noalign{\smallskip}
 Event & $T_{0}$ [TBJD] & $P$ [d] & $a/R_{\star}$ & $R_{\rm p}/R_{\star}$ &  $i_{\rm p}$ [$^{\circ}$] & $\sigma$ [ppm] \\
        \noalign{\smallskip}
        \hline
 1\tablefootmark{a} & $1342.22\pm0.03$ & $30\pm6$ & $40\pm8$ & $0.028\pm0.006$ & $89.28\pm0.45$ & - \\
\hline
 1 & $1342.2247^{+0.0005}_{-0.0004}$ & $17^{+13}_{-8}$ & $25^{+18}_{-11}$ & $0.036^{+0.0014}_{-0.0012}$ & $88.6^{+0.9}_{-1.2}$ & 388 \\
2 & $2040.0063^{+0.0005}_{-0.0004}$ & $23^{+7}_{-5}$ & $27\pm6$ & $0.0420^{+0.0007}_{-0.0009}$ & $88.28^{+0.3}_{-0.5}$ & 678 \\
3 & $2058.8672\pm0.0004$ & $21^{+46}_{-12}$ & $37^{+78}_{-21}$ & $0.0359^{+0.0009}_{-0.0005}$ & $89.35^{+0.52}_{-1.03}$ & 646 \\
\hline
1 & $1342.2243^{+0.0004}_{-0.0006}$ & $18.8574^{+0.0016}_{-0.0008}$ & $31.3^{+0.4}_{-0.8}$ & $0.0393^{+0.0005}_{-0.0004}$ & $89.04^{+0.09}_{-0.06}$ & 355 \\
2 & $2040.0062\pm0.0003$ & $18.8588^{+0.00012}_{-0.00014}$ & $23.8^{+1.5}_{-1.3}$ & $0.0444\pm0.0007$ & $88.2\pm0.2$ & 650 \\
3 & $2058.8671\pm0.0004$ & $18.8592^{+0.0002}_{-0.0003}$ & $31.5^{+0.2}_{-2.1}$ & $0.0409\pm0.0004$ & $89.03^{+0.07}_{-0.20}$ & 610 \\
    \hline
    \end{tabular}
    \tablefoot{
    \tablefoottext{a}{Fit parameters for Event 1 obtained by \cite{plavchan2020Natur}.}
    }
  \end{center}
  \end{table*}

\end{appendix}

\end{document}